\documentclass[aps,amsfonts,amssymb,epsfig,pre]{revtex4}

\usepackage{epsfig}
\usepackage{amsmath}
\usepackage{graphicx} 
\DeclareMathAlphabet{\mathpzc}{OT1}{pzc}{m}{it}
\draft
\input{epsf}

\begin{document}

\title{Phase behavior of a symmetrical binary fluid mixture}

\author{J\"urgen K\"ofinger$^1$, Nigel B. Wilding$^2$, and Gerhard Kahl$^3$}

\affiliation{$^1$ Faculty of Physics, University of Vienna,
  Boltzmanngasse 5, A-1090 Wien, Austria \\ $^2$ Department of
  Physics, University of Bath, Bath BA2 4LP, United Kingdom \\ $^3$
  Center for Computational Materials Science and Institut f\"ur
  Theoretische Physik, TU Wien, Wiedner Hauptstra{\ss}e 8-10, A-1040
  Wien, Austria}

\date{\today}

\begin{abstract}

We have investigated the phase behaviour of a symmetrical binary fluid
mixture for the situation where the chemical potentials $\mu_1$ and
$\mu_2$ of the two species differ. Attention is focused on the set of
interparticle interaction strengths for which, when $\mu_1=\mu_2$, the
phase diagram exhibits both a liquid-vapor critical point and a
tricritical point. The corresponding phase behaviour for the case
$\mu_1\ne\mu_2$ is investigated via integral-equation theory
calculations within the mean spherical approximation (MSA), and grand
canonical Monte Carlo (GCMC) simulations. We find that two possible
subtypes of phase behaviour can occur, these being distinguished by the
relationship between the critical lines in the full phase diagram in the
space of temperature, density, and concentration. We present the
detailed form of the phase diagram for both subtypes and compare with
the results from GCMC simulations, finding good overall agreement.
The scenario via which one subtype evolves into the other, is also
studied, revealing interesting features.

\end{abstract}

\pacs{~}

\maketitle

\linespread{1.4}

\section{Introduction}

As one proceeds from a simple single component fluid to a binary
mixture, the complexity of the phase behaviour increases dramatically.
The range of possible behavior was first summarized on a qualitative
level more than 25 years ago in the seminal study of van Konynenburg
and Scott \cite{Scott}.  Using a simple mean-field (MF) ansatz for the
equation-of-state, these authors found that in binary mixtures the
additional thermodynamic degree of freedom in the form of the
concentrations of the two species, $c_1$ and $c_2$, with $c \equiv c_1
= 1 - c_2$ (or, alternatively, the difference in the chemical
potentials of the two species, $\Delta \mu = \mu_1 - \mu_2$),
considerably widens the spectrum of possible features \cite{Row82}.
Among these are upper and lower critical points, tricritical points,
critical end points (CEPs), quadruple (i.e., four phase) points, lines
of critical points, or triple lines. The details of which features
actually emerge in practice for a given model, and in which order
(i.e., the phase diagram topology), were found to depend in a
non-trivial manner on the precise combination of interparticle
interaction parameters, i.e., three energy parameters $\epsilon_{ij}$
and three length scales $\sigma_{ij}$ (with $i,j=1,2$), together with
parameters that characterize the range of the attractive interactions.

Today, more than a quarter of a century later, both liquid state
theories \cite{Han86} and computer simulations \cite{Fre02,Lan00} offer
standards of accuracy that greatly surpass those obtainable from
MF-based theories. It therefore seems appropriate to reconsider the
schematic overview of phase diagram topologies provided by van
Konynenburg and Scott \cite{Scott}, at a more {\it quantitative} level. 
The present contribution aims to take a step in this direction. However,
owing to the high-dimensional space of interaction parameters, one is
necessarily restricted to a particular model. In the interests of
numerical tractability, we have chosen to consider a {\em symmetrical}
binary mixture. In such a system, the interactions between particles of
like species are equal, i.e., $\Phi_{11}(r) = \Phi_{22}(r) = \Phi(r)$,
while the interactions between unlike species are related via
$\Phi_{12}(r) = \delta \Phi(r)$. Thus for a prescribed form of
interaction potential, $\Phi(r)$, the phase behaviour is controlled by
the single parameter $\delta$. Since for the case $\delta >1$, such a
system exhibits only a simple liquid-vapour (LV) transition, we focus
here on the more interesting case of $\delta < 1$, where the competition
between a LV on the one hand, and a demixing transition (into phases of
generally different concentrations), gives rise to a
rather rich phase behaviour. For the reasons explained below, we will
focus on a mixture where $\Phi(r)$ takes the form of a hard-core plus
attractive Yukawa (HCY) interaction; nevertheless we anticipate that any
other similar forms of potential (such as Lennard-Jones or square-well
potentials) will lead to results qualitatively similar to those we
present.

In recent years, numerous authors have investigated the phase diagram of
symmetrical binary mixtures using liquid state theories (of varying
degrees of sophistication), as well as computer simulations
\cite{Pan88,Rec93,Cac93,Gre94,deM95,Wil97,Cac98,Wil98,
Kah01,Ant02,Wil03,Woy03,Sch03,Pin03,Sch04,Sch05,Sch05a,Pas01}. However,
these studies have been confined for the large part to the special case
of equimolar concentrations of the two species (i.e., for $\Delta \mu =
0$). The resulting picture of the phase behaviour shows four distinct
topologies (or archetypes) of phase diagram as $\delta$ is varied. 
Starting at $\delta = 1$, and in order of decreasing $\delta$, these are
distinguishable as follows: in the first type, the critical line that
characterizes the demixing transition (the so-called $\lambda$-line)
intersects the LV coexistence line at liquid densities and at
temperatures well below the LV critical temperature, forming a critical
end point (CEP); in the second type, the $\lambda$-line approaches LV
coexistence again from the liquid side, but slightly below the LV
critical temperature, terminating in a separate tricritical point which
itself marks the end of a first order transition between mixed and
demixed liquid phases; in the third type the $\lambda$-line intersects
the LV critical point directly, the two combining to form a tricritical
point. Finally in the fourth type the $\lambda$-line intersects the low
density branch of the LV coexistence curve, forming a CEP. For clarity
we summarize in Table~\ref{tab:systemparameters} the various notations
that have been introduced in literature to classify the phase diagrams
of binary symmetrical mixtures for $\Delta \mu = 0$. A schematic
presentation of the first three types can be found in Fig.~1 of
\cite{Wil98}, while the fourth type is shown in Fig.~1 of \cite{Sch05a}.
We note that in analytical calculations the precise value of $\delta$
marking the boundaries between two successive types for a particular
interaction potential, depends to some extent on the method used for the
calculation.

For the more general case of non-equimolar concentrations (i.e.,
allowing for $\Delta \mu \ne 0$), relatively few investigations of the phase
behaviour can be found in literature \cite{Pin03,Ant02,Woy06}.
This fact undoubtedly reflects the obstacles encountered by both
simulation and theoretical approaches in dealing with the increased
complexity of the full phase diagram spanned by temperature $T$,
pressure $p$, {\em and} the chemical potential difference $\Delta
\mu$. Probably the first study to provide a rough idea of the
complexity of the full phase diagram was that of Antonyevich {\it
et al.} \cite{Ant02}, based on conventional integral-equation
techniques. It was followed shortly afterwards by the remarkable study
of Pini {\it et al.} \cite{Pin03}, who used the highly accurate
Hierarchical Reference Theory (HRT) \cite{Par95}, complemented by MF
calculations, to provide the first insight into the complex topology
of the {\it full} phase diagram of binary symmetrical mixtures for
$\Delta \mu \ne 0$. However, the high computational costs of HRT did
not permit a systematic discussion of the topology of the phase
diagram.

In this work we present a systematic investigation of the full phase
diagram for one of the archetypes of phase diagram outlined above,
namely that exhibiting both a LV critical point and a first order
liquid-liquid transition terminating at a tricritical point
\cite{Koef06}. Our investigations are based on the mean spherical
approximation (MSA) and grand canonical Monte Carlo (GCMC) simulations.
We have chosen MSA for the simple reason that for a mixture of HCY
fluids it provides closed (i.e., semi-analytic) expressions for the
structural and the thermodynamic properties \cite{Hoy78}; these concern
in particular the pressure, $p$, and the chemical potentials, $\mu_1$
and $\mu_2$. The formalism for a general mixture of HCYs is summarized
in \cite{Arr87}. Using then {\tt MATHEMATICA}\texttrademark~\cite{Math},
the solution of the coexistence equations is considerably less involved
than it is the case for integral-equation theories that have to be
solved fully numerically \cite{Han86}. Although MSA is less accurate (in
particular in the critical regions) than, e.g., advanced liquid state
theories, such as HRT \cite{Par95} or the self-consistent
Ornstein-Zernike approximation (SCOZA) \cite{Pin98,Sch03,Sch05}, it is
nevertheless able to provide results of reasonable accuracy; this will
also be corroborated in the present contribution. GCMC, when used in
combination with extended sampling and histogram reweighting techniques
provides a powerful means of studying fluid phase equilibria with a
precision which is limited only by statistical errors
\cite{Wil95,Wil98}. We extend its use here to the case of a chemical
potential difference field, thus providing a benchmark against which to
test the MSA results.

Our work makes progress on two fronts. First, we demonstrate that
liquid state theories and computer simulations have reached levels of
accuracy and efficiency that the full phase diagram [i.e., in $(T,
\rho, c)$- or in $(T, p, \Delta \mu)$-space] of a symmetrical binary
fluid mixture can be determined on a semi-quantitative level. We note
that even a few years ago such an aim was considered to be too
ambitious \cite{Pin03}. Second, we show that our simple model exhibits
a wider spectrum of topologies of phase behaviour than one might have
expected from previous investigations, in particular from those that
were restricted to the equimolar case. Specifically, we confirm that
for the archetype of phase behaviour considered (whose associated
domain of $\delta$ is defined on the basis of the $\Delta \mu =0$
phase diagram topology), two distinct subtypes emerge when one
considers $\Delta \mu \ne 0$, as already suggested by Pini {\it et
al.}  \cite{Pin03} on the basis of selected isothermal cuts through
the phase diagram. Our exposition of the {\em full} phase diagram
sheds further light on the nature of the subtypes, by showing that
their topologies are defined by the location of four coexistence
surfaces (each of which carries a critical line), which intersect in
four triple lines.  Some of these triple lines do not originate at the
$\lambda$-line at equal concentrations.  We consider the locus of the
triple lines to be the key quantities that distinguish the two
subtypes of phase behaviour. We discuss in detail the transition
scenario between the two subtypes, which has not yet been documented in
literature and which reveals new and interesting details about the
critical behaviour of the system.  The existence of the two subtypes
is confirmed by the results of the GCMC simulation studies which,
while being less comprehensive than the theoretical ones, demonstrate that
simulation can be predictive regarding quite subtle features of the
phase diagram. 

Although a binary symmetrical mixture might at first sight seem to be a
system of purely academic interest, arguments have been put forward in
literature that they are able to reflect features of realistic systems:
van Konynenburg and Scott \cite{Scott} note that `the only real binary
systems in which such symmetry occurs are mixtures of $d, l$-optical
isomers \cite{Sco77}', while Woywod and Schoen \cite{Woy06} find that
the phase behaviour of $^3$He-$^4$He mixtures contains many features of
the topology of the phase diagram of binary symmetrical mixtures.
Furthermore, as we shall discuss later, the symmetrical mixture has
direct relevance to some classes of spin fluid models.

The paper is organized as follows. In Sec.~\ref{sec:model}, we
introduce our model, a symmetrical binary fluid mixture in which
particles interact via the HCY potential. The MSA integral-equation
strategy is then outlined, as is the grand canonical simulation
methodology. Sec.~\ref{sec:results} represents in turn our MSA and
GCMC results for the phase behaviour of the system for non-zero
chemical potential difference field. Finally in Sec.~\ref{sec:concs},
we attempt to set our results within their wider context, and assess
the outlook for future work.

\section{Model and methods}

\subsection{The model}
\label{sec:model}

We consider a binary symmetrical mixture where the particles interact
via the HCY potential

\begin{equation}                                  \label{phi_tot}
\Phi_{ij}(r) = \left\{ \begin{array} {l@{~~~~~~~~}l}
                             \infty  & r \le \sigma \\
                             -\epsilon_{ij} \sigma \frac{1}{r}
                             \exp[-z (r - \sigma)]    & r > \sigma \\
                             \end{array}
                             \right. .
\end{equation}
Here the $\epsilon_{ij}$ are the contact values of the potentials,
$\sigma$ is the hard-core diameter, and $z$ is the inverse screening
length, assigned the value $z \sigma = 1.8$ in this work. Further the
system is characterized by a temperature $T$ [with $\beta = (k_{\rm B}
T)^{-1}$] and a number density $\rho =N/V$, $N$ being the total number
of particles in a volume $V$. Defining $c$ as the concentration of
species 1, we introduce partial number densities $\rho_1 = c \rho$ and
$\rho_2 = (1-c) \rho$; in the expressions below we use $x_1=c$ and
$x_2=1-c$. Standard reduced units will be used throughout the paper.

\subsection{Mean spherical approximation}

The solution of the Ornstein-Zernike equations \cite{Han86} along with
the MSA closure relation, i.e.,

\begin{equation}  \begin{array}{lcll} c_{ij}(r) & = & - \beta
\Phi_{ij}(r) & r > \sigma \\ g_{ij}(r) & = & 0 & r \le \sigma \\ 
\end{array} 
\end{equation}  
can be carried out for this particular system to a large extent
analytically \cite{Hoy78}. Here the $c_{ij}(r)$ and the $g_{ij}(r)$ are
the direct correlation functions and the pair distribution functions,
respectively. Our solution strategy is based on the general formalism
presented by Arrieta {\it et al.}  \cite{Arr87} for the MSA solution of
multi-component HCY systems. These authors have derived a set of
coupled, non-linear equations for the initially unknown coefficients that
characterize the correlation functions $c_{ij}(r)$. These equations
require a fully numerical solution (see below); for details we refer the
interested reader to ref.~\cite{Arr87}. Once these coefficients
are obtained, essentially all thermodynamic properties can be calculated via simple
algebraic expressions. Below we present only those that we require for
the present contribution.

Within the MSA the thermodynamic properties derived via the energy
route are usually more accurate than via other routes
\cite{Lomba}. In the following, we list expressions for the excess
(over the hard-sphere (HS) reference system) pressure and chemical
potentials. The corresponding properties of the reference system are
given by the Carnahan-Mansoori-Starling-Leland (CMSL) equation of
state \cite{CMSL}. With $p_E = p_E^{\rm CMSL} + \Delta p_E$ and $\mu_i
= \mu_i^{\rm CMSL} + \Delta \mu_i$ ($i=1,2$), these quantities are
given by

\begin{equation}
\frac{\beta p^{\mathrm{CMSL}}}{\rho}=
\frac{1}{1-\eta}+\frac{18(\xi_1 \xi_2+\xi_2^2)}{\pi\rho(1-\eta)^2}+
\frac{6 \eta \xi_2^3}{\pi\rho(1-\eta)^3}
\end{equation}
 
\begin{equation}
\frac{\beta \Delta p_E}{\rho}=
\frac{\pi}{3}\rho\sum_{ij} x_i x_j \sigma^3 \{ [g_{ij}(\sigma)]^2-
[g_{ij}^0(\sigma)]^2 \} + J
\end{equation}
 
\begin{align}
\beta \mu_i^{\mathrm{CMSL}} \nonumber
& = \ln \left(\frac{\rho_i}{1-\eta}\right)+
\frac{\frac{\pi}{6}\rho\sigma^3+3\sigma\xi_2+3\sigma^2\xi_1}{1-\eta}+\\  
\nonumber
& +\frac{3\sigma^3\xi_1\xi_2+3\sigma^3\xi_2^2}{(1-\eta)^2}+
\frac{3 \sigma^2\xi_2^2}{\eta(1-\eta)^2}+\\ \nonumber
& +\frac{3\sigma^2\xi_2^2\ln(1-\eta)}{\eta^2}-
\frac{(\sigma^3\xi_2^3)(2-5\eta+\eta^2)}{\eta^2(1-\eta)^3}+\\ 
& -\frac{2\sigma^3\xi_2^3 \ln(1-\eta)}{\eta^3}
\end{align}

\begin{equation}
\beta \Delta  \mu_i=
-\frac{2\pi}{z}\sum_j\rho_j K_{ij} G_{ij} - 
\frac{1}{2} \sum_j \rho_j[\tilde c_{ij}(0)-\tilde c_{ij}^0(0)]
\end{equation} 
where $\xi_j = \pi/6 (\rho \sigma^j)$, $j=1,2,3$; in particular
$\xi_3 = \eta$, the packing fraction. Further parameters that appear
in these relations (such as $J$, $G_{ij}$, and $K_{ij}$) are {\it a
priori} unknown quantities, which are obtained from the numerical
solution of the non-linear equations mentioned above.  The
$g_{ij}(\sigma)$ and the $g^0_{ij}(\sigma)$ are the values of the pair
distribution functions of the HCY and of the HS reference system at
contact. Finally, $\tilde c_{ij}(0)$ and $\tilde c_{ij}^0(0)$ are the
zero-components of the Fourier-transform of the direct correlation
functions for the HCY and the HS system. Details of the numerical 
solution of the MSA which leads to the full phase diagram will be 
presented in the Appendix.

\subsection{Computer simulations}

Many features of the simulation techniques employed in the present study
have previously been detailed elsewhere \cite{Wil95,Wil97,Wil98}.
Accordingly, we confine the description of our methodology to its
essentials, except where necessary to detail a new aspect.

GCMC simulations were performed using a standard Metropolis algorithm
\cite{ALLEN,Wil98}. The simulation scheme comprises two types of
operations:

\begin{enumerate}
\item Particle insertions and deletions. 
\item Particle identity transformations: $1\to 2$, $2\to 1$. 
\end{enumerate}
Since particle positions are sampled implicitly via the random
particle transfer step, no additional particle translation moves are
required. Indeed it is the sampling of density fluctuations which
represent the bottleneck for this problem, and the incorporation of
particle translations does nothing to improve sampling efficiency.

The simulations were performed using the potential of
eq.~(\ref{phi_tot}), truncated at $r_c=3.0\sigma$. A mean-field long
range correction for the effects of this truncation was applied in the
usual manner~\cite{Fre02}. To minimize the overheads associated with
identification of neighbor interactions, we employed a cell
decomposition scheme \cite{ALLEN}. This involves partitioning the
periodic simulation space of volume $L^3$ into $l^3$ cubic cells, each
of linear dimension of the interaction range, i.e., $L/l=3.0$. We chose
to study system sizes corresponding to $l=4$, $l=7$, and $l=9$. The
smallest system size was used to trace phase boundaries away from
critical points; additionally, the larger ones were employed near
criticality to facilitate crude finite-size scaling estimates of
critical point locations. At each state point, equilibration periods of
up to $2\times10^3$ MCS (Monte Carlo sweeps, where one sweep comprises
$l^3$ attempted transfers and identity swaps) were used. Sampling
frequencies were $500$ MCS for the $l=4$ system size rising to $2000$
MCS for the $l=9$ system. The total number of samples drawn for each
state point studied was about $10^6$.

In this work we explore the parameter space spanned by the variables
$(\mu_1,\Delta\mu,T,\delta)$. To accomplish this, without having to
perform a very large number of simulations, we employed the histogram
reweighting technique \cite{FERRENBERG}. Use of this technique permits
histograms obtained at one set of model parameters to be reweighted to
yield estimates corresponding to another set of model parameters. To
enable simultaneous reweighting in all four fields
$(\mu_1,\Delta\mu,T,\delta)$, one must
sample the conjugate observables $(\rho_1,\rho_1-\rho_2,u,u_d)$, with
$\rho_1=N_1/V$, $\rho_2=N_2/V$, $u$ the configurational energy
density, and $u_d$ that part of $u$ associated with interactions
between {\em dissimilar} particle species. These quantities were
accumulated in the form of a list during the simulations, and
histograms of their forms were created and reweighted via
post-processing of the entries in the list.

Within a region of first order coexistence, a standard grand canonical
ensemble scheme is severely hampered by the large free energy barrier
separating the two coexisting phases. This barrier leads to
metastability effects and prohibitively long correlation times.  To
circumvent this difficulty, we have employed the multicanonical
preweighting method \cite{BERG} which enhances sampling of the
interfacial configurations of intrinsically low probability. This is
achieved by incorporating a suitably chosen weight function in the
GCMC update probabilities. The weights are subsequently `folded out'
from the sampled distributions of observables to yield the correct
Boltzmann distributed quantities. Use of this method permits the
direct measurement of the distribution of observables at first order
phase transitions, even when these distributions span many decades of
probability. Details concerning the implementation of the techniques
can be found in references \cite{BERG,Wil01}.

Phase boundaries were traced in the space of ($\mu_1,\Delta\mu,T$) by
applying the equal peak weight criterion \cite{Borgs} to the
distribution of the appropriate fluctuating order parameter, whether
this be the concentration or the overall density. Critical point
parameters were estimated using a crude version of the finite-size
scaling (FSS) analysis described in ref. \cite{Wil95}. The analysis
involves scanning the parameter space until the observed probability
distribution of the fluctuating order parameter matched the
independently known universal fixed point form appropriate to the Ising
universality class in the FSS limit \cite{Bloete}.

\section{Results}
\label{sec:results}
\subsection{MSA results}

In the following we present the MSA results for the two subtypes of
phase diagram, denoted $\alpha$ and $\beta$, that we have distinguished.
For each subtype we discuss its form first in the space of the physical
fields, $(T, p, \Delta \mu)$, and then in a mixed field-density space,
$(T, \rho, c)$. Although representations in field space are somewhat
less conventional than those in a mixed space of fields and densities,
they do afford a compact display of results because two-phase
coexistence occurs on surfaces. Taken together both representations
provide a full picture of the characteristic features of the respective
topologies. We note that as a consequence of the symmetry of the
underlying model the field space representation is symmetric with
respect to the plane $\Delta \mu = 0$, while the mixed space
representation is symmetric with respect to $c = 1/2$.

\subsubsection{The type $\alpha$ phase diagram}

\begin{figure}[t]
\epsfig{angle=0, file=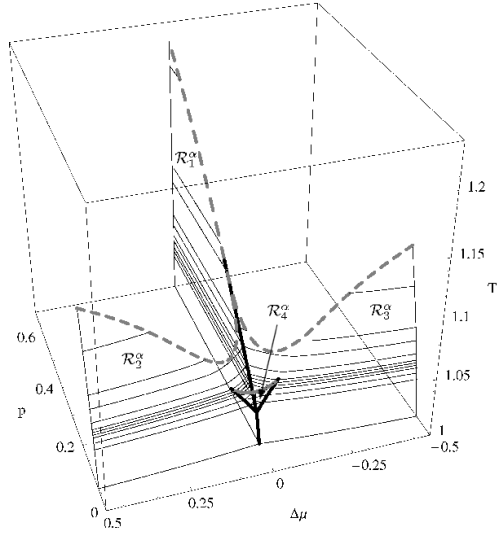, width=6.5cm, clip}
\caption{MSA results for the type $\alpha$ phase diagram of the binary
   symmetrical mixture ($\delta =0.67$) considered in this study in
   $(T, p, \Delta \mu)$-space. Symbols: thin full lines -- isothermal
   coexistence lines, grey full thick line -- critical line
   $\mathfrak{c}_4^\alpha$ passing through the LV critical point, grey
   dashed thick lines -- critical lines ($\mathfrak{c}_1^\alpha$,
   $\mathfrak{c}_2^\alpha$, and $\mathfrak{c}_3^\alpha$) passing
   through the tricritical point, black thick
   lines -- triple lines $\mathfrak{t}_i^\alpha$, $i=1, \dots, 4$.}
\label{phd_alpha_field}
\end{figure}

Here we consider the phase behaviour of subtype $\alpha$. Motivated by
the results of Pini {\it et al.} \cite{Pin03}, we have obtained the
phase behaviour using MSA for $\delta = 0.67$. We present first, in
Fig.~\ref{phd_alpha_field}, the phase diagram in $(T, p, \Delta
\mu)$-space. Four coexistence surfaces are distinguishable, ${\cal
R}_i^\alpha$, $i=1,\dots, 4$, along each of which two phases coexist.
${\cal R}_1^\alpha$ lies in the $(\Delta \mu=0)$-plane and represents
the demixing surface. ${\cal R}_2^\alpha$ and ${\cal R}_3^\alpha$ are
symmetrically related and describe neither pure demixing transitions
nor pure liquid-vapour (LV) transitions. Finally, along ${\cal
R}_4^\alpha$ there is (predominantly) LV coexistence. The lines of
intersection between ${\cal R}_2^\alpha$ and ${\cal R}_3^\alpha$ with
${\cal R}_4^\alpha$ are symmetrically related triple lines,
$\mathfrak{t}_2^\alpha$ and $\mathfrak{t}_3^\alpha$.  On the high
temperature side they terminate in critical end points (CEPs), which
are connected by a critical line, $\mathfrak{c}_4^\alpha$, marking the
upper edge of surface ${\cal R}_4^\alpha$. This critical line passes
through the LV critical point of the field-free case, located at
$\Delta \mu = 0$. The surfaces ${\cal R}_2^\alpha$, ${\cal
R}_3^\alpha$, and ${\cal R}_4^\alpha$ thus form a pocket delimited by
the critical line $\mathfrak{c}_4^\alpha$ and the triple lines
$\mathfrak{t} _2^\alpha$ and $\mathfrak{t}_3^\alpha$.  In this pocket
a homogeneous, mixed fluid is stable as was observed previously in
ref.~\cite{Pin03}. Two further triple lines, $\mathfrak{t}_1^\alpha$
and $\mathfrak{t}_4^\alpha$, can be found in the equimolar plane of
${\cal R}_1^\alpha$: they are the intersection lines of the two
symmetrically related surfaces, ${\cal R}_2^\alpha$ and ${\cal
R}_3^\alpha$, with the demixing surface ${\cal R}_1^\alpha$. The fact
that these three surfaces intersect in a line rather than in a point
is a consequence of the symmetry of the system. All triple lines meet
in the only quadruple point of the system. The equimolar triple line
$\mathfrak{t}_1^\alpha$ intersects at high temperatures the
tricritical point of the system. At this point three critical lines,
$\mathfrak{c}_1^\alpha$, $\mathfrak{c}_2^\alpha$, and
$\mathfrak{c}_3^\alpha$, originate: in the equimolar plane of ${\cal
R}_1^\alpha$ lies the critical line $\mathfrak{c}_1^\alpha$ that
characterizes the demixing transition (the so-called $\lambda$-line)
and the two symmetrically related critical lines
$\mathfrak{c}_2^\alpha$ and $\mathfrak{c}_3^\alpha$ that specify the
mixing-demixing transition, and which tend (for $|\Delta \mu| \to
\infty$) towards the critical points of the respective pure
components. If we focus on the isotherms that are marked in
Fig.~\ref{phd_alpha_field} we recover at a quantitative level the
scenario depicted in Figs. 3(b) to (d) of ref. \cite{Pin03}.

\begin{figure}[t]
\epsfig{angle=0,file=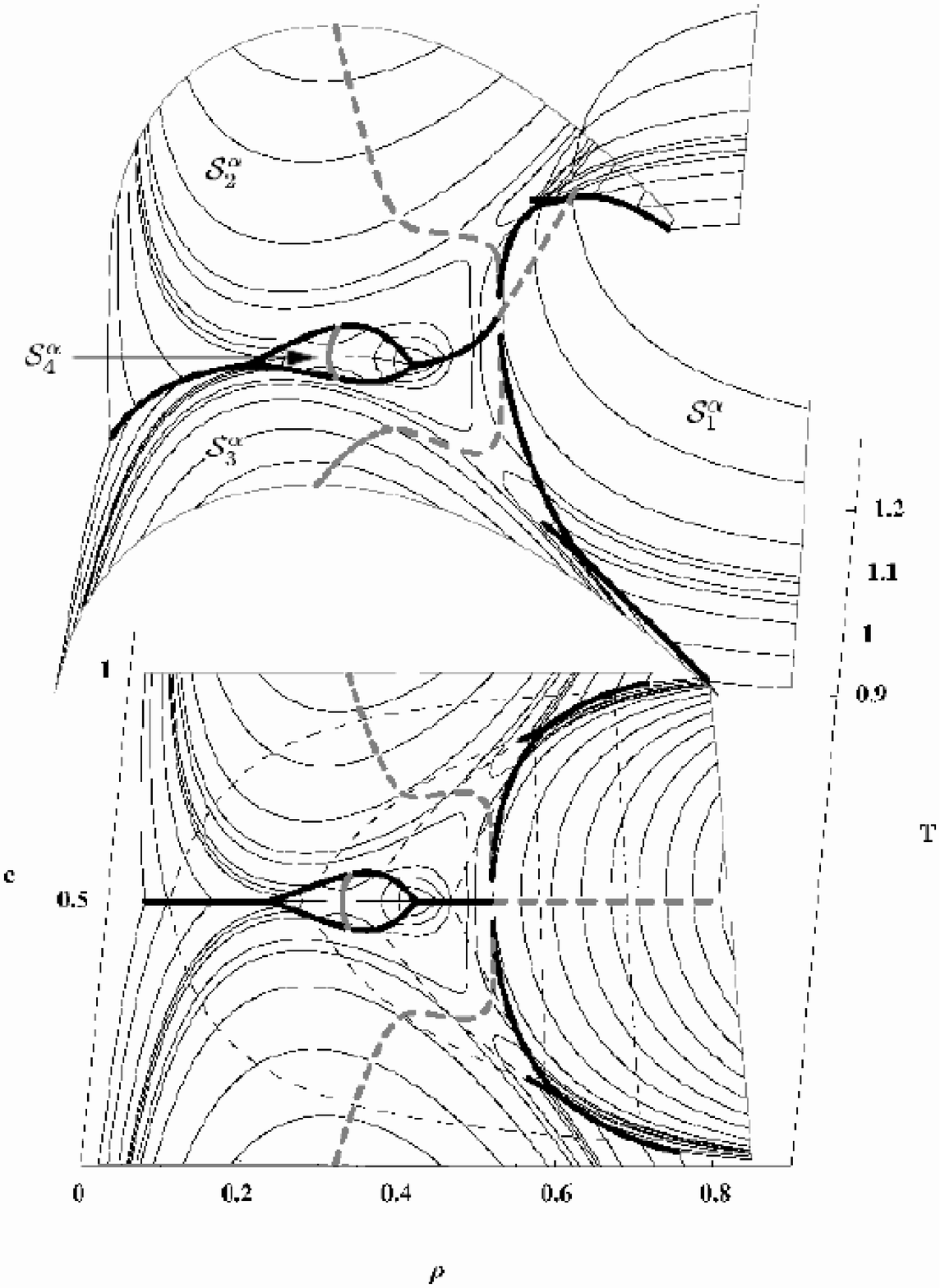, width=7cm, clip}
\caption{MSA results for the type $\alpha$ phase diagram of the binary
   symmetrical mixture ($\delta=0.67$) considered in the study in $(T,
   \rho, c)$-space and its projection onto the $(\rho,
   c)$-plane. Symbols: thin full lines -- isothermal coexistence
   lines, dashed thin lines -- tie lines, grey full thick line --
   critical line $\mathfrak{c}_4^\alpha$ passing through the LV
   critical point of field-free case, grey dashed thick lines --
   critical lines ($\mathfrak{c}_1^\alpha$, $\mathfrak{c}_2^\alpha$,
   and $\mathfrak{c}_3^\alpha$) passing through the tricritical point
   of the field-free case, black thick lines -- triple lines
   $\mathfrak{t}_i^\alpha$, $i=1, \dots, 4$.}
\label{phd_alpha_mixed}
\end{figure}

Turning next to the mixed field-density representation for subtype
$\alpha$ [i.e., in $(T, \rho, c)$-space], this is displayed in
Fig.~\ref{phd_alpha_mixed}. Four distinct coexistence surfaces, ${\cal
S}_i^\alpha$, are observable, corresponding to the respective ${\cal
R}_i^\alpha$, $i=1 \dots, 4$. In this representation, however, the
surfaces separate `two-phase' regions (below) from regions of
homogeneous phases (above).  Again, surfaces intersect along triple
lines now represented by three lines (branches). Each surface exhibits
a critical line, $\mathfrak{c}_i^\alpha$. $\mathcal{S}_1^\alpha$ is
symmetrical with respect to the ($c=1/2$)-plane and its critical line,
$\mathfrak{c}_1^\alpha$, is the $\lambda$-line. On
$\mathcal{S}_1^\alpha$, two symmetrical high-density phases ($\rho
\gtrsim 0.5$), specified by $(\rho, c)$ and $(\rho, 1-c)$, are in
coexistence. The two (symmetrically related) coexistence surfaces,
$\mathcal{S}_2^\alpha$ and $\mathcal{S}_3^\alpha$, are encountered for
low and intermediate densities $(\rho \lesssim 0.6)$; the order
parameter that characterizes this transition is a linear combination
of the density and concentration differences of the two coexisting
phases.  Coming from high temperatures the $\lambda$-line bifurcates
at the tricritical point into two critical lines,
$\mathfrak{c}_2^\alpha$ and $\mathfrak{c}_3^\alpha$ (belonging to
$\mathcal{S}_2^\alpha$ and $\mathcal{S}_3^\alpha$), which pass through
minima (i.e., `double critical points') and head for the respective LV
critical points of the pure phases. The fourth coexistence surface,
$\mathcal{S}_4^\alpha$, is predominantly LV-like in character and is
located in between $\mathcal{S}_2^\alpha$ and $\mathcal{S}_3^\alpha$.
Its critical line passes through the LV critical point of the
field-free mixture.  The surface as a whole is delimited by the {\sf
four} intermediate-density branches of the triple lines
$\mathfrak{t}_2^\alpha$ and $\mathfrak{t}_3^\alpha$ which form a
lens-shaped `loop'. Tie-lines starting at these enclosing triple lines
connect a vapour and a liquid phase of approximately equal
concentrations with a liquid phase of higher density and different
concentration [see $(\rho, c)$-projection of
Fig.~\ref{phd_alpha_mixed}, where for each triple line a set of tie
lines was selected]. The latter states form the high-density branch of
the triple lines $\mathfrak{t}_2^\alpha$ and $\mathfrak{t}_3^\alpha$,
that are located in the `valley(s)' formed by $\mathcal{S}_2^\alpha$
(or $\mathcal{S}_3^\alpha$) and $\mathcal{S}_1^\alpha$.
In ($T, \rho, c)$-space the quadruple point can
be localized as follows: its intermediate density representations are
located on the end points of the lens-shaped `loop', while the other
two, being symmetrically related, are the bifurcation points of the
triple lines $\mathfrak{t}_2^\alpha$ and $\mathfrak{t}_3^\alpha$.  As
the critical line $\mathfrak{c}_4^\alpha$ of $\mathcal{S}_4^\alpha$
approaches the bounding triple lines $\mathfrak{t}_2^\alpha$ and
$\mathfrak{t}_3^\alpha$, the coexisting vapour and liquid phases
become critical. Since these are simultaneously in equilibrium with a
non-critical phase (the so-called spectator phase) located at the end
point of the high-density branch of the triple lines, this point is a
CEP. The region of a homogeneous, mixed fluid at intermediate
densities found in \cite{Pin03} and discussed above in field space,
is bounded by the three coexistence surfaces $\mathcal{S}_2^\alpha$,
$\mathcal{S}_3^\alpha$, and $\mathcal{S}_4^\alpha$.

\begin{figure}[t]
\epsfig{angle=0, file=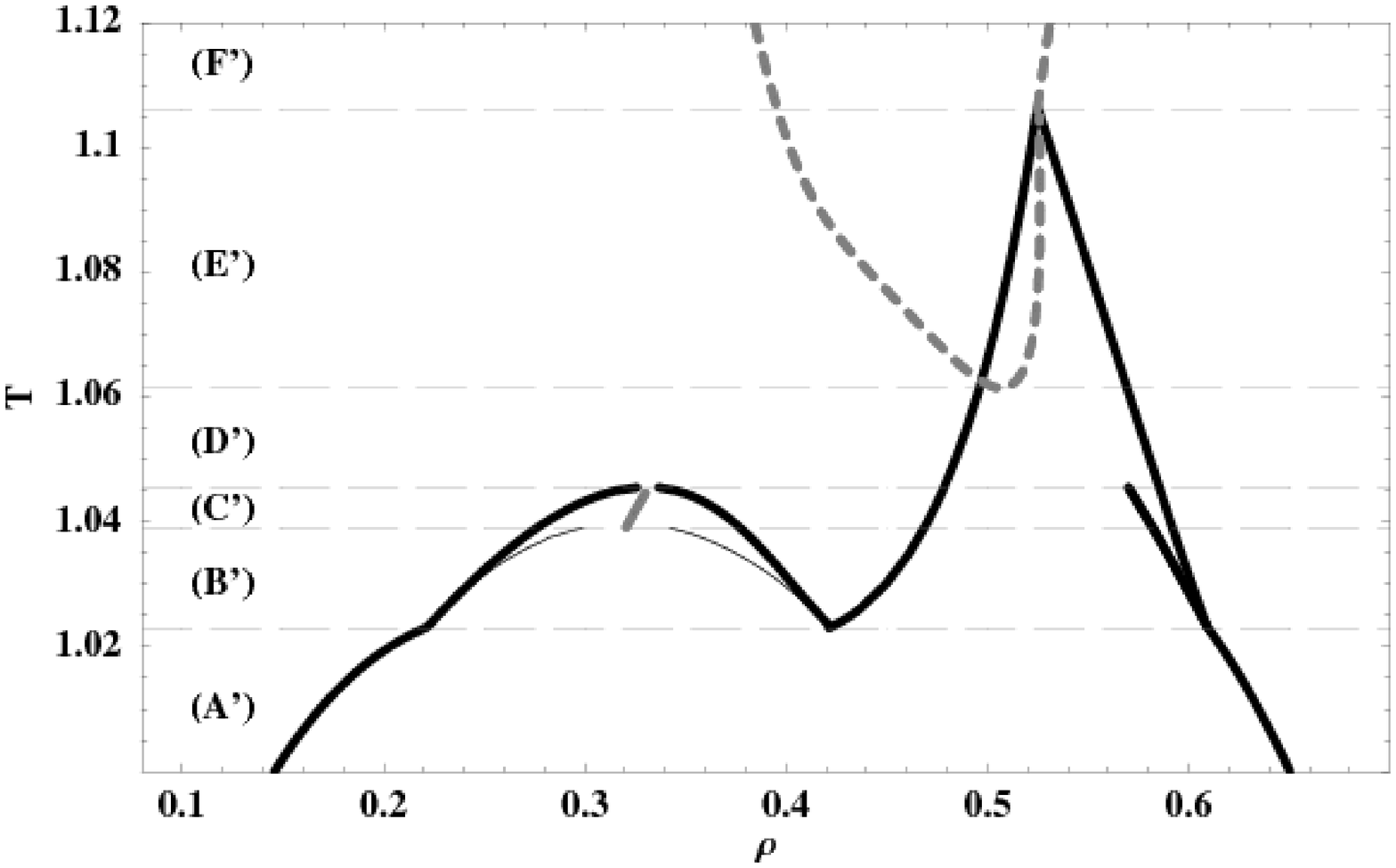, width=6.5cm, clip}
\caption{MSA results for the type $\alpha$ phase diagram of the binary
   symmetrical mixture ($\delta=0.67$) considered in the study,
   projected onto the $(T,\rho)$-plane. Dashed thin lines separate
   ranges A' to F' (see text). Line symbols see Figs.
   \ref{phd_alpha_field} and \ref{phd_alpha_mixed}; in addition: thin
   line --  azeotropic (i.e., field-free coexistence) line.}
\label{phd_alpha_T_rho}
\end{figure}

\begin{figure}[t]
\epsfig{angle=0, file=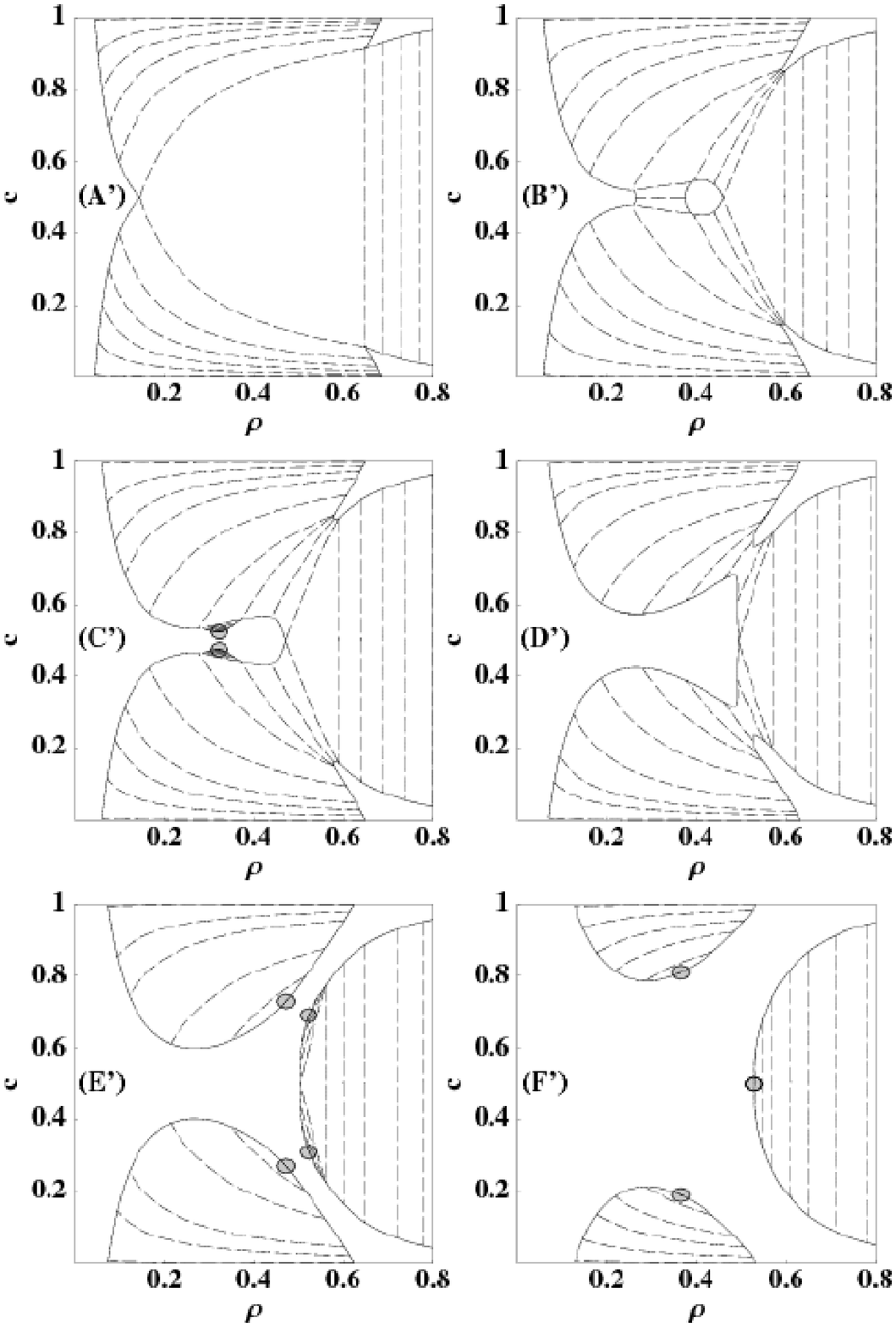, width=6.5cm, clip}
\caption{Six different panels display MSA results for the type
   $\alpha$ phase diagram of the binary symmetrical mixture
   ($\delta=0.67$) considered in the study, projected onto the $(\rho,
   c)$-plane.  The six panels labeled A' to F' show isothermal cuts at
   the temperatures $T=\{1., 1.035, 1.041, 1.06, 1.07, 1.15\}$ and
   correspond to the six temperature ranges indicated in
   Fig. \ref{phd_alpha_T_rho}.}
\label{phd_alpha_rho_c}
\end{figure}

We close with a more qualitative discussion of the type $\alpha$ phase
diagram which, nevertheless, might be rather instructive. In Fig.~
\ref{phd_alpha_T_rho} we show the projection of the phase diagram in
$(T, \rho, c)$-space onto the $(T, \rho)$-plane. In this representation
we have introduced six temperature ranges (labeled A' to F'); they are
separated by five isotherms which are characterized by particular
temperatures of the system. In Fig. \ref{phd_alpha_rho_c} we show
representative isothermal cuts taken from each of these temperature
ranges in the $(\rho, c)$-plane. These figures complement those of
Fig.~2 in ref. \cite{Pin03} in the sense that they present a more
comprehensive sequence of possible scenarios of coexistence regions
encountered in the type $\alpha$ phase diagram. However, we note that
within subtype $\alpha$, other scenarios in the isothermal cuts can also
be observed. This is, for instance, the case when the LV critical point
lies {\it above} the temperature that characterizes the minimum in the
symmetrically related critical lines $\mathfrak{c}_2^\alpha$ and
$\mathfrak{c}_3^\alpha$ that pass through the equimolar tricritical
point. Hence, scenarios different to the ones presented here can be
encountered. In range A', i.e., at temperatures below the quadruple
point, regions of homogeneous phases are only found at low densities
(extending over all concentrations), and in small areas at high
densities (with concentrations close to 0 or 1). As indicated by the tie
lines, two-phase coexistence is generally observed; only at $c=1/2$ and
low densities and at the symmetrically related corners of the
homogeneous high density phases does three phase coexistence occur. At
the temperature of the quadruple point, an infinitesimal amount of a
fourth homogeneous phase emerges at $c=1/2$ and $\rho \simeq 0.45$. As
we proceed to range B', this point extends to a small region where a
homogeneous phase exists (see the discussion above); as can be seen from
the tie lines, its three corners are points of three phase coexistence.
Furthermore the other regions of homogeneous phases encountered already
in range A' grow both in density and concentration. 

By further increasing the temperature we pass through the equimolar LV
critical point and reach range C'. Now a narrow passage has opened up
between the homogeneous mixed fluid ($c \sim 1/2$ and $\rho \sim 0.5$)
and the homogeneous low density phases, that becomes broader with
increasing temperature. It is characterized by two critical points
located on the symmetrically related coexistence regions. At some
temperature these become CEPs and we enter the range D'. Here the
scenario is dominated by three huge, interconnected coexistence regions:
one is symmetrical with respect to $c=1/2$ at high density and two are
symmetrically related (one for $c \gtrsim 0.6$, the other one for $c
\lesssim 0.4$). In the range D', the coexistence regions are nearly
entirely two-phase in character; three-phase coexistence is only
observed at intermediate and high densities. The next transition
temperature is characterized by the minimum of the symmetrically related
critical lines $\mathfrak{c}_2^\alpha$ and $\mathfrak{c}_3^\alpha$ that
originate at the tricritical point and head towards the critical points
of the pure components. As we pass this temperature, we enter range E'
and the connections between the three coexistence regions break,
leaving them entirely disconnected. Four critical points can now be
observed: two of them are located on the symmetrically related
coexistence regions and two others (symmetrically related), can be found
on the demixing coexistence region. In small areas, close to these
regions, three-phase coexistence is observed. As we finally pass through
the tricritical point of the system we enter range F', where the three
coexistence regions are well-separated and the phase diagram is
dominated by huge areas of homogeneous mixtures. The coexistence regions
now show exclusively two-phase coexistence. The two critical points of
the demixing coexistence region have merged at the tricritical point to
a single critical point (located on the $\lambda$-line).

\subsubsection{The type $\beta$ phase diagram}

\begin{figure}[b]
\epsfig{angle=0, file=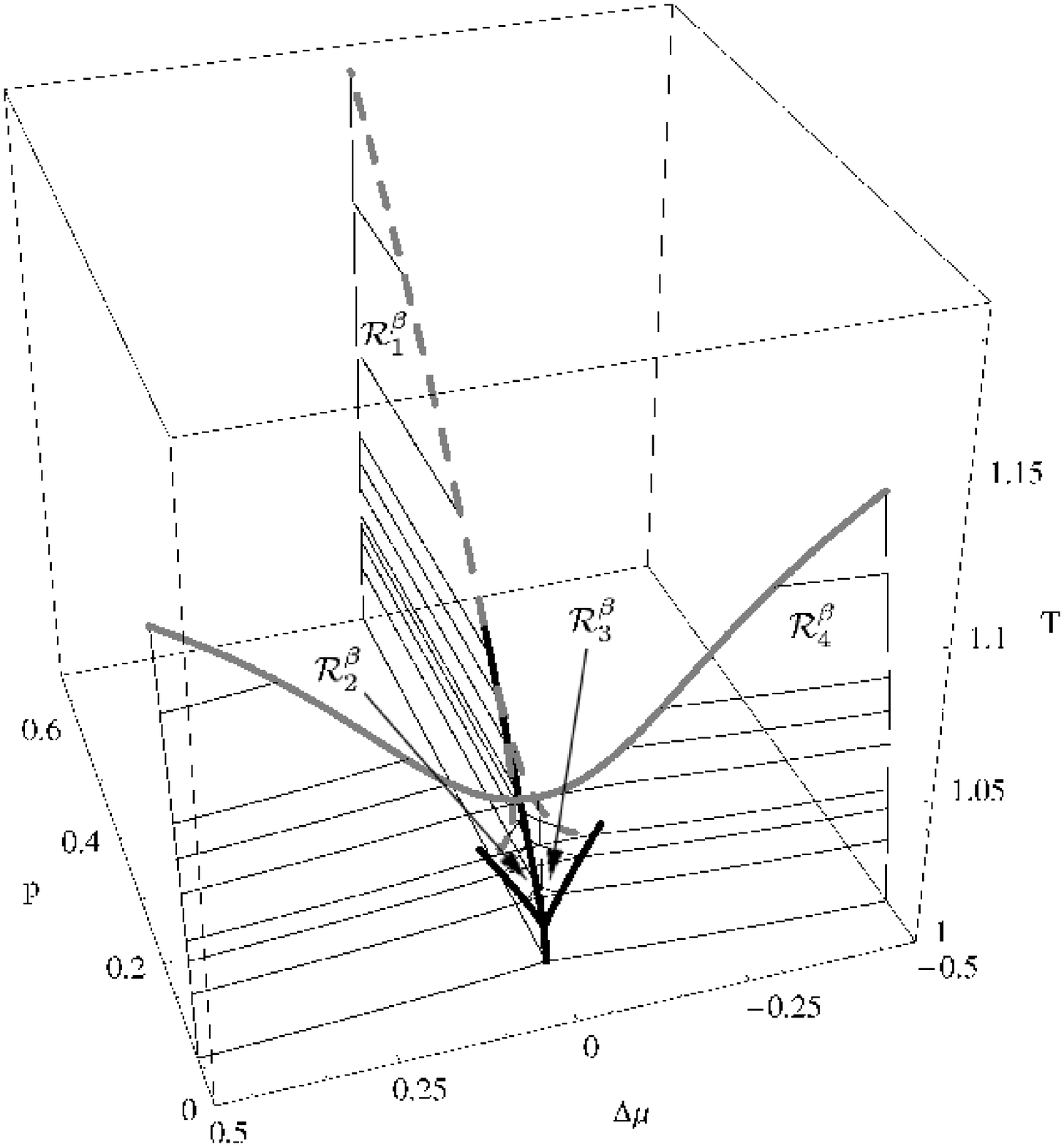, width=6.5cm, clip}
\caption{MSA results for the type $\beta$ phase diagram of the binary
   symmetrical mixture ($\delta =0.69$) considered in this study in
   $(T, p, \Delta \mu)$-space. Symbols: thin full lines -- isothermal
   coexistence lines, grey full thick line -- critical line
   $\mathfrak{c}_4^\beta$ passing through the LV critical point, grey
   dashed thick lines -- critical lines ($\mathfrak{c}_1^\beta$,
   $\mathfrak{c}_2^\beta$, and $\mathfrak{c}_3^\beta$) passing through
   the tricritical point, black thick lines --
   triple lines $\mathfrak{t}_i^\beta$, $i=1, \dots, 4$.}
\label{phd_beta_field}
\end{figure}

In the type $\beta$ phase diagram, which we have found to occur at
$\delta = 0.69$, the phase behaviour is distinct from that of type
$\alpha$. In the $(T, p, \Delta \mu)$-space representation of Fig. 
\ref{phd_beta_field} we can again clearly identify four coexistence
surfaces ${\cal R}_i^\beta$, $i=1,\dots, 4$. However, the manner of
their intersection engenders a different topology.  The triple line
$\mathfrak{t}_4^\beta$ is now the intersection of the surfaces ${\cal
R}_1^\beta$ and ${\cal R}_4^\beta$.  Intersection of ${\cal R}_2^\beta$
and ${\cal R}_3^\beta$ with ${\cal R}_1^\beta$ leads again to the triple
line $\mathfrak{t}_1^\beta$, in the $(\Delta \mu=0)$-plane, which
terminates at high temperatures at the tricritical point. From there,
(and similarly to type $\alpha$), three critical lines originate: the
$\lambda$-line (i.e., the critical line $\mathfrak{c}_1^\beta$ of ${\cal
R}_1^\beta$) continues to higher temperatures, while two symmetrically
related critical lines, $\mathfrak{c}_2^\beta$ and
$\mathfrak{c}_3^\beta$, head off to lower temperatures. These terminate in
two CEPs where they also meet two symmetrically related triple lines,
$\mathfrak{t}_2^\beta$ and $\mathfrak{t}_3^\beta$.  Along these lines
${\cal R}_2^\beta$ and ${\cal R}_3^\beta$ intersect with ${\cal
R}_4^\beta$ (which is a predominantly LV coexistence surface) and which
dominates the low pressure region and covers the whole $\Delta
\mu$-range.  The intersection of $\mathfrak{t}_2^\beta$ and
$\mathfrak{t}_3^\beta$ with the aforementioned triple lines
$\mathfrak{t}_1^\beta$ and $\mathfrak{t}_4^\beta$, in the ($\Delta
\mu=0$)-plane, gives rise to the only quadruple point of the system.
${\cal R}_4^\beta$ is delimited at high temperatures by a further
critical line, $\mathfrak{c}_4^\beta$, which connects the critical
points of the pure components (i.e., for $|\Delta \mu| \to \infty$) with
the LV critical point of the mixture at $\Delta \mu = 0$.

\begin{figure}[t]
\epsfig{angle=0,file=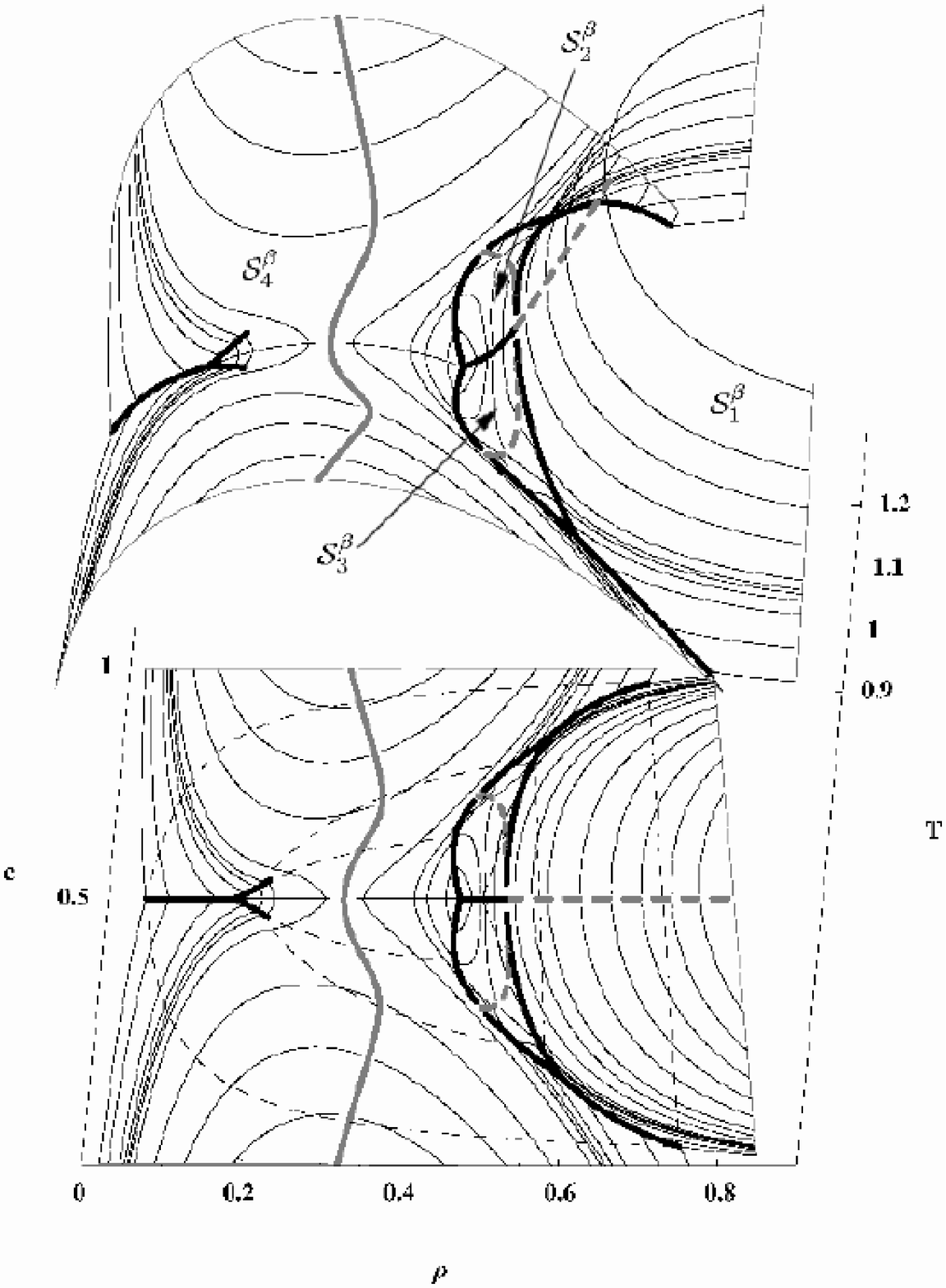, width=7cm, clip}
\caption{MSA results for the type $\beta$ phase diagram of the binary
   symmetrical mixture ($\delta=0.69$) considered in the study in $(T,
   \rho, c)$-space and its projection onto the $(\rho,
   c)$-plane. Symbols: thin full lines -- isothermal coexistence
   lines, dashed thin lines -- tie lines, grey full thick line --
   critical line $\mathfrak{c}_4^\beta$ passing through the LV
   critical point of field-free case, grey dashed thick lines --
   critical lines ($\mathfrak{c}_1^\beta$, $\mathfrak{c}_2^\beta$, and
   $\mathfrak{c}_3^\beta$) passing through the tricritical point of
   the field-free case, black thick lines -- triple lines
   $\mathfrak{t}_i^\beta$, $i=1, \dots, 4$.}
\label{phd_beta_mixed}
\end{figure}

Again considering the mixed field-density representation in ($T, \rho,
c$)-space, the distinction between type $\beta$ and type $\alpha$ phase
diagrams is clear. The type $\beta$ case is depicted in Fig.
\ref{phd_beta_mixed}. Four coexistence surfaces ${\cal S}_i^\beta$ with
accompanying critical lines $\mathfrak{c}_i^\beta$ can be identified;
they correspond to the respective ${\cal R}_i^\beta$. The $\lambda$-line
($\mathfrak{c}_1^\beta$), being the critical line of the symmetrical
demixing surface ${\cal S}_1^\beta$, bifurcates at the tricritical point
into two critical lines, $\mathfrak{c}_2^\beta$ and
$\mathfrak{c}_3^\beta$, which traverse the (symmetrically related)
surfaces $\mathcal{S}_2^\beta$ and $\mathcal{S}_3^\beta$.  Passing
through minima (i.e., `double critical points') these critical lines
terminate in CEPs located at the symmetrically related high density
branches of the triple lines, $\mathfrak{t}_2^\beta$ and
$\mathfrak{t}_3^\beta$.  These triple lines are the intersection of
$\mathcal{S}_2^\beta$ and $\mathcal{S}_3^\beta$ with coexistence surface
$\mathcal{S}_4^\beta$ that encompasses the entire concentration range.
At these CEPs, two high-density phases become critical while in
coexistence with the spectator phase located at the end points of the
low-density branches of the triple lines. The critical line
$\mathfrak{c}_4^\beta$ on $\mathcal{S}_4^\beta$ passes through the
equimolar LV critical point and connects the critical points of the
respective pure components; in contrast to type $\alpha$ it is now
completely detached from the $\lambda$-line. In subtype $\beta$ the
triple lines also show a distinctively different behaviour than their
equivalents in subtype $\alpha$.  As can be seen from the representative
tie lines (all connecting coexistence points on triple lines) shown in
Fig.~\ref{phd_beta_mixed}, a single low-density phase (with $c\simeq
1/2$) is connected via tie lines to two high-density phases.  In contrast
to type $\alpha$, the high and the intermediate density triple lines,
$\mathfrak{t}_1^\beta$, $\mathfrak{t}_2^\beta$, and
$\mathfrak{t}_3^\beta$, merge to form two symmetrically related `loops',
while the low density branches terminate as the spectator phase of the
CEPs.  Again, for $\rho \sim 0.5$ a region of a homogeneous, mixed fluid
is encountered, which is enclosed by the coexistence surfaces
$\mathcal{S}_2^\beta$, $\mathcal{S}_3^\beta$, and $\mathcal{S}_4^\beta$
(see also ref.~\cite{Pin03}).  At the quadruple point, the low
density branch of $\mathfrak{t}_4^\beta$ bifurcates in the low density
branches of $\mathfrak{t}_2^\beta$ and $\mathfrak{t}_3^\beta$, whereas
the high density branches of $\mathfrak{t}_4^\beta$ bifurcate in the
high density branches of $\mathfrak{t}_2^\beta$ (and symmetrically
related $\mathfrak{t}_3^\beta$) and $\mathfrak{t}_1^\beta$.  The fourth
representation of the quadruple point is located in the `valley' of
mixed fluid, where the intermediate branches of $\mathfrak{t}_2^\beta$,
$\mathfrak{t}_3^\beta$, and $\mathfrak{t}_1^\beta$ intersect. 

\begin{figure}[t]
\epsfig{angle=0, file=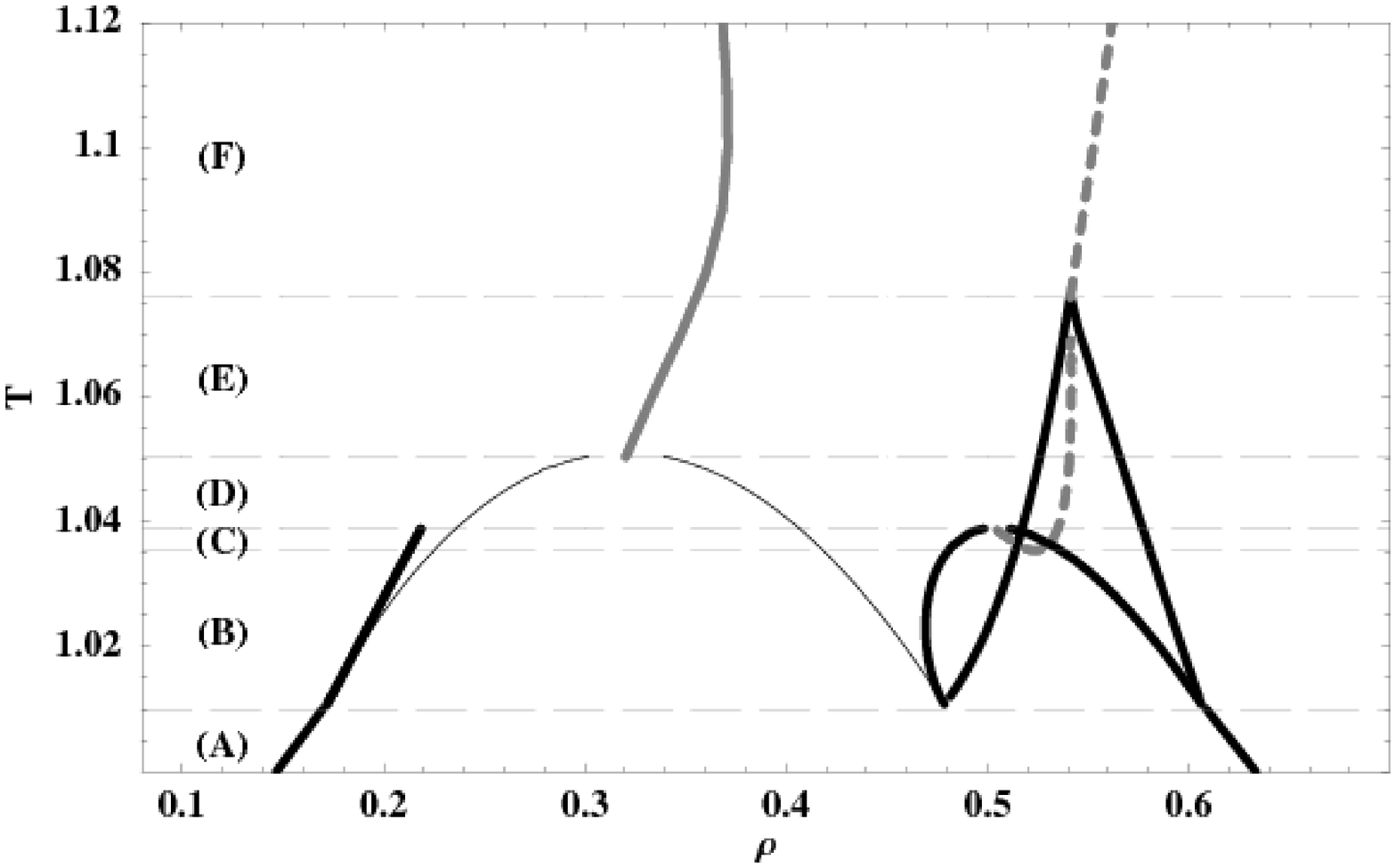, width=6.5cm, clip}
\caption{MSA results for the type $\beta$ phase diagram of the binary
   symmetrical mixture ($\delta=0.69$) considered in the study
   projected onto the $(T,\rho)$-plane. Dashed thin lines separate
   ranges A to F (see text). Line symbols see Figs.
   \protect\ref{phd_beta_field} and \ref{phd_beta_mixed}; in addition: thin
   line -- azeotropic (i.e., field-free coexistence) line.}
\label{phd_beta_T_rho}
\end{figure}

\begin{figure}[t]
\epsfig{angle=0, file=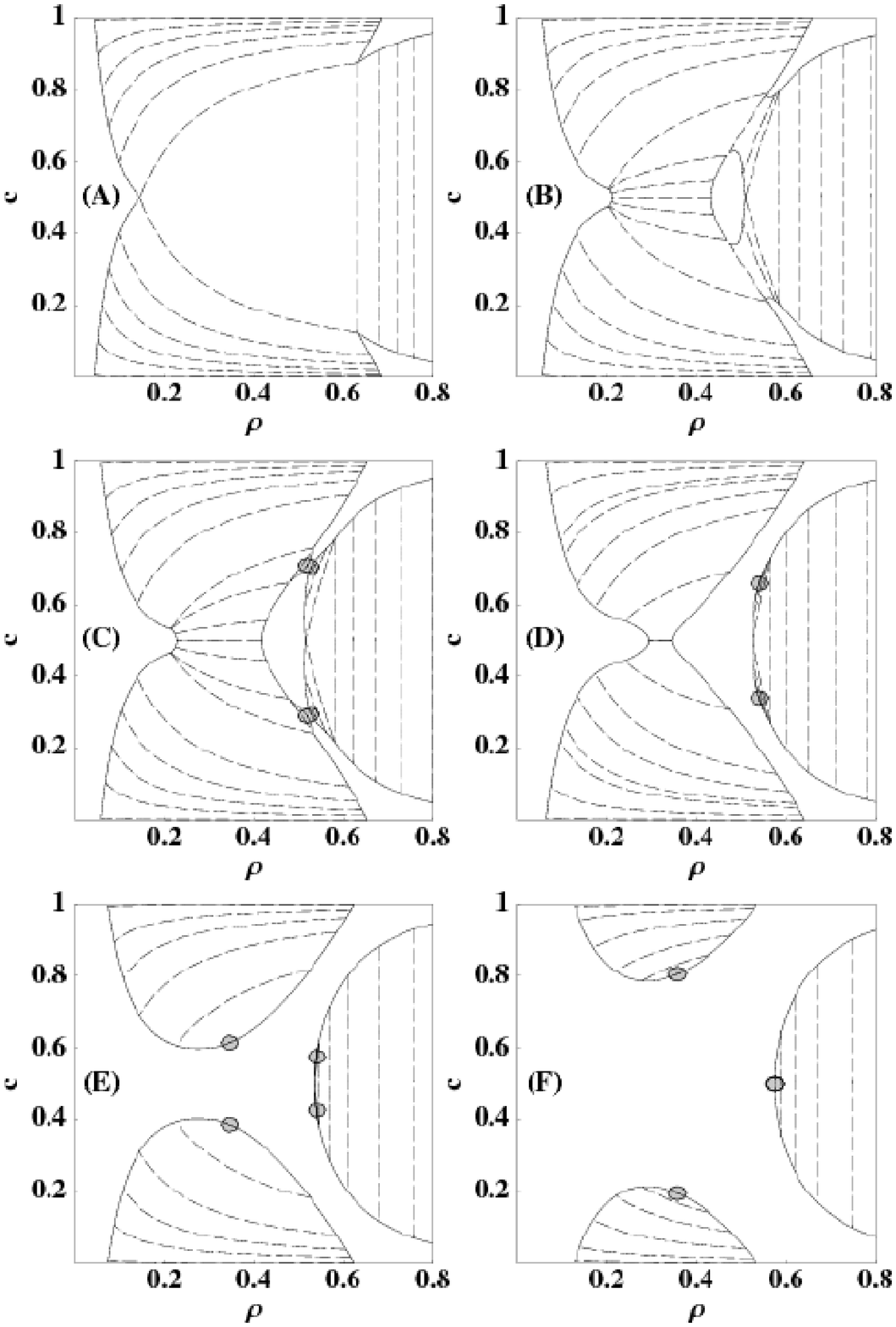, width=6.5cm, clip}
\caption{Six different panels display MSA results for the type $\beta$
   phase diagram of the binary symmetrical mixture ($\delta=0.69$)
   considered in the study, projected onto the $(\rho, c)$-plane.  The
   six panels labeled A to F show isothermal cuts at the temperatures
   $T=\{1, 1.03, 1.036, 1.05, 1.07, 1.15\}$ and correspond to the six
   temperature ranges indicated in Fig.  \ref{phd_beta_T_rho}.}
\label{phd_beta_rho_c}
\end{figure}

Again we close the discussion of the type $\beta$ phase diagram with a
more qualitative discussion. In Fig. \ref{phd_beta_T_rho} we show
projections of characteristic lines of the phase diagram onto the $(T,
\rho)$-plane, defining six temperature ranges, labeled A to F. In Fig.
\ref{phd_beta_rho_c} we show isothermal cuts for each of these
temperature ranges.  In the type $\beta$ phase diagram the scenarios in
the temperature ranges A and B (which are again separated by the
quadruple point temperature) correspond on a qualitative level exactly
to the corresponding ones in type $\alpha$. In particular, in range B we
again encounter three large interconnected coexistence regions with a
small area of a homogeneous, mixed fluid for $c \sim 1/2$ and $\rho \sim
0.5$.  The minimum on the symmetrically related critical lines,
$\mathfrak{c}_2^\beta$ and $\mathfrak{c}_3^\beta$, on ${\cal S}_2^\beta$
and ${\cal S}_3^\beta$ (mentioned above) define the temperature where we
enter range C: in contrast to type $\alpha$, the three large coexistence
regions break up at $\rho \gtrsim 0.5$, splitting off the demixing
coexistence region and leaving the other two connected. The
symmetrically related passages are characterized by pairs of critical
points (located on each of the two coexistence regions in a 3d
representation in $(T, \rho, c)$-space), and connected by symmetrically
related critical lines $\mathfrak{c}_2^\beta$ and $\mathfrak{c}_3^\beta$
that originate in the tricritical point and terminate at symmetrically
related CEPs.  This CEP-temperature marks the limit between ranges C and
D. In range D only the symmetrically related critical points on the
demixing surface remain; simultaneously, the connecting bridge between
the large coexistence regions that extend over all concentrations
becomes narrower. Passing the equimolar LV critical point (and thus entering
range E) this connection breaks up leading to two symmetrically
related coexistence regions that are entirely detached from the demixing
region, and on each of which a pair of symmetrically related critical
points emerges. Finally, above the tricritical point of the system
(i.e., in the range F) the three coexistence regions are by now well
separated, each being characterized by a critical point, which
corresponds exactly to the scenario depicted for the range F' in subtype
$\alpha$. Again we note that within subtype $\beta$, slightly different
scenarios in the isothermal cuts can be observed. This is, for instance,
the case if the equimolar LV critical temperature is lower than the
temperature of the minimum in the symmetrically related triple lines,
$\mathfrak{t}_2^\beta$ and $\mathfrak{t}_3^\beta$.

\subsubsection{The transition scenario}
\label{sec:transition}

Of course it is of particular interest how the transition from one
subtype to the other takes place. We start from subtype $\alpha$ and
increase the parameter $\delta$.  In the $(T, p, \Delta
\mu)$-representation the transition scenario between the two subtypes
can be traced easily. The characteristic pocket formed by ${\cal
R}_2^\alpha$, ${\cal R}_3^\alpha$, and ${\cal R}_4^\alpha$ becomes
larger by extending the limiting triple line branches,
$\mathfrak{t}_2^\alpha$ and $\mathfrak{t}_3^\alpha$ (and thus the
connecting critical line $\mathfrak{c}_4^\alpha$) in both directions.
Simultaneously the symmetrically related critical lines
$\mathfrak{c}_2^\alpha$ and $\mathfrak{c}_3^\alpha$ that emerge from the
tricritical point are shifted to lower temperatures. At the transition
value $\tilde \delta$, these critical lines have reached the CEPs,
thereby forming two additional, symmetrically related tricritical
points.  As a consequence the coexistence surface $\mathcal{R}_2^\alpha$
is subdivided by the triple line $\mathfrak{t}_2^\mathrm{tr}$ (the
former $\mathfrak{t}_2^\alpha$), into two sub-surfaces,
$\mathcal{R}_{2'}^\mathrm{tr}$ and $\mathcal{R}_{2''}^\mathrm{tr}$. The
same happens with the symmetrically related surface
$\mathcal{R}_{3}^\alpha$.  A three sided pocket is now formed by
$\mathcal{R}_{2''}^\mathrm{tr}$, $\mathcal{R}_{3''}^\mathrm{tr}$, and
$\mathcal{R}_{4}^\mathrm{tr}$ (the former $\mathcal{R}_{4}^\alpha$).
From below it is bounded by the quadruple point, its edges are the
triple lines, the upper three corners are the tricritical points which,
in turn, are connected by sections of the critical lines (see below).
Upon further increasing $\delta$, $\mathcal{R}_{4}^\mathrm{tr}$ merges
with $\mathcal{R}_{2'}^\mathrm{tr}$ and $\mathcal{R}_{3'}^\mathrm{tr}$
forming thus $\mathcal{R}_{4}^\beta$ while
$\mathcal{R}_{2''}^\mathrm{tr}$ ($\mathcal{R}_{3''}^\mathrm{tr}$)
becomes $\mathcal{R}_{2}^\beta$ ($\mathcal{R}_{3}^\beta$).  The surface
$\mathcal{R}_{1}^\alpha$ is not affected by the transition and
corresponds to $\mathcal{R}_{1}^\beta$ .  In a similar manner the
critical line $\mathfrak{c}_2^\alpha$ (and symmetrically related
$\mathfrak{c}_3^\alpha$) is subdivided at the transition in two sections
$\mathfrak{c}_{2'}^\mathrm{tr}$ and $\mathfrak{c}_{2''}^\mathrm{tr}$,
where $\mathfrak{c}_{2'}^\mathrm{tr}$ is delimited by the field-free
tricritical point and one of the newly formed tricritical points.
$\mathfrak{c}_{4}^\mathrm{tr}$ is limited by the newly formed
tricritical points.  Upon further increasing $\delta$,
$\mathfrak{c}_{4}^\mathrm{tr}$, $\mathfrak{c}_{2''}^\mathrm{tr}$, and
$\mathfrak{c}_{3''}^\mathrm{tr}$ merge to form $\mathfrak{c}_{4}^\beta$.
$\mathfrak{c}_{2'}^\mathrm{tr}$ ($\mathfrak{c}_{3'}^\mathrm{tr}$)
becomes $\mathfrak{c}_{2}^\beta$ ($\mathfrak{c}_{3}^\beta$) and are
delimited by CEPs and the field-free tricritical point.

\begin{figure}[b]
\epsfig{angle=0, file=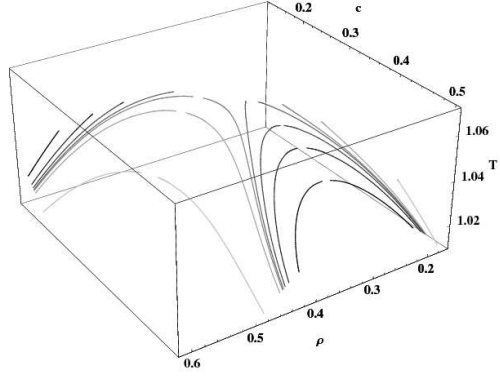, width=6.5cm, clip}
\caption{View of the triple lines $\mathfrak{t}_3^\alpha$ and
   $\mathfrak{t}_3^\beta$ from the equimolar plane ($c=1/2$) in $(T,
   \rho, c)$-space for seven different values of $\delta= \{0.67,
   0.675, 0.677, 0.678, 0.6785, 0.68, 0.69\}$ denoted in gray scales
   from light gray to black.}
\label{transition}
\end{figure}

To discuss the transition scenario in the $(T, \rho, c)$-representation
we again start from type $\alpha$ and increase $\delta$.  Then gradually
the high- and the intermediate-density branches of the triple lines
$\mathfrak{t}_2^\alpha$ and $\mathfrak{t}_3^\alpha$ lengthen until they
merge when the CEP located on the intermediate-density branch meets the
end point of the high-density branch, thereby forming a tricritical
point (see Fig.~\ref{transition}).  Upon further increase of $\delta$, a
high-density branch (with a CEP) and a low-density branch of the triple
line detach, thus resulting in the topology of the $\beta$-subtype.
Concomitant with this metamorphosis of the triple lines, we observe a
related development  in the critical line $\mathfrak{c}_4^\alpha$ that
passes through the LV critical point in subtype $\alpha$: with
increasing $\delta$, this line lengthens in both directions until it
meets (at the crossover between the subtypes) the critical lines of the
surfaces $\mathcal{S}_2^\alpha$ and $\mathcal{S}_3^\alpha$. As $\delta$
is further increased, a new critical line forms connecting the critical
points of the pure phases, passing through the equimolar LV critical
point so that the CEPs remain with the newly formed triple lines at
higher densities.

Within the MSA the transition occurs for $\tilde \delta_{\rm MSA}
= 0.678(0)$.  We derive this estimate by following the development of
the triple lines $\mathfrak{t}_2^\alpha$ and $\mathfrak{t}_3^\alpha$
to $\mathfrak{t}_2^\beta$ and $\mathfrak{t}_3^\beta$ with increasing
$\delta$.  In both subtypes, two of the three branches of the triple
lines mentioned above meet each other in CEPs, whereas at the
transition all three branches meet in a tricritical
point. Fig.~\ref{transition} shows a selection of the triple lines
$\mathfrak{t}_3^\alpha$ and $\mathfrak{t}_3^\beta$ for different
values of $\delta$.  

The transition scenario (although not explicitly identified
there) is nicely depicted on several occasions in \cite{Pin03}. Within
the MF framework, where the transition occurs at $\tilde \delta_{\rm
MF} = 0.65338$, the above mentioned loop of critical lines is nicely
depicted in Fig. 1(c). Figs. 10 and 12 show projections of the HRT
$(T, \rho, c)$ phase diagram onto the $(\rho, c)$-plane at various
temperatures for $\delta = 0.67$, while in Fig. 11 isothermal cuts
through the $(T, p, \Delta \mu)$ phase diagram are shown. This
$\delta$-value is close to the HRT-transition value, $\tilde
\delta_{\rm HRT}$.

\subsubsection{Relationship to previous theoretical work}

The two subtypes that we discuss here have already been classified by
van Konynenburg and Scott \cite{Scott}: type $\alpha$ corresponds to
`sym. III-A$^\star$' while type $\beta$ to type `sym. II-A$^\star$'.
Several years later Antonevych {\it et al.} \cite{Ant02} pointed out
that the complex topology of the {\it full} phase diagram of a binary
symmetrical mixture originates from the complex interplay of the LV
transition and the demixing transition (see Figs. 1 to 4 in this
contribution). However, most of their considerations were carried out
on a {\it qualitative} level (see section III.B of ref.~\cite{Ant02}).

Shortly afterwards, Pini {\it et al.} \cite{Pin03} presented their
remarkable HRT and MF theoretical study of the full phase diagram of
binary symmetrical mixtures, discovering several interesting features
which they presented in the form of isothermal cuts through the phase
diagrams in $(T, p, \Delta \mu)$- and in $(T, \rho, c)$-space. The two
subtypes that we have discussed here in detailed were identified 
in their contribution as the two alternative ways in which the region
of the homogeneous fluid occurring at intermediate densities (see
above) is formed from the intersection of the coexistence surfaces.
Additionally, the authors pointed out that in what we term subtype
$\alpha$, the critical temperature at $\Delta \mu = 0$ is always lower
than the tricritical temperature, while in subtype $\beta$ no such
ordering relation holds. While Pini {\it et al.} focus in their
contribution on a classification of the topology via the critical
lines, we consider a discussion of the phase diagrams via the location
of the triple lines to be more enlightening. In fact, our 3d
representations both in the space of the physical fields as well as in
the mixed field-density space helped us to clarify the following
points: (i) from these representations we obtained an unambiguous
identification of two distinct subtypes (confirming also
the conclusions drawn in \cite{Pin03}); (ii) in the 3d plots, the
`double critical points' reported in \cite{Pin03} are identifiable
as local minima in the critical lines; (iii) we find previously
unreported triple lines;  (iv) the transition scenario between the
two subtypes could be discussed and clarified in detail (see
sec. \ref{sec:transition}).  Comparison of our isothermal cuts
with those presented by Pini {\it et al.} facilitates the link between
our subtypes and the topological changes identified in \cite{Pin03}:
in their Figs. 2, 3, 6, and 7 a type $\alpha$ behaviour can be
identified, while Figs. 8 and 9 display a $\beta$ topology; Figs. 10
to 12 depict a close-to-transition scenario (see also
sec.~\ref{sec:transition}).

Finally, Woywod and Schoen \cite{Woy06}  have very recently studied a
lattice model using a MF density functional theory, to calculate the
complete phase diagram of a binary fluid mixture of equally sized
particles. The conclusions that the authors could draw from their
investigations are relevant also for the present study: they present
arguments (that back up earlier conclusions \cite{TRICRIT}) which
preclude the existence of tricritical points in {\it general} binary
mixtures, but allow them under certain symmetry conditions which are
fulfilled by symmetrical mixtures, but can also be found in particular
asymmetric mixtures such as those investigated in \cite{Woy06}.  Their
investigations of the binary {\it symmetrical} mixture is focused on the
shape of the phase diagram in field space for selected values of a
parameter $\Lambda$, that characterizes the relative strength of the
mixed interaction (cf. \cite{Scott}), thus following essentially the
sequence of systems summarized in Table~\ref{tab:systemparameters}.
However, the existence of the two subtypes $\alpha$ and $\beta$ is not
mentioned at all in this contribution. However, due to the fact that
Woywod and Schoen could calculate thermodynamic potentials via an
essentially  closed expression, they were able to pin down exactly the
transition scenario discussed in sec. \ref{sec:transition}. In
particular in their Fig.~5 this transition in the $(T, \mu, \Delta
\mu)$-space is clearly depicted, showing explicitly the three
tricritical points, the quadruple point, and the connecting critical and
triple lines that occur for this particular system. We point out that
for the semi-analytical approach used in the present contribution within
the MSA framework, such an exact localization of the transition is
beyond reach.

\subsection{Simulation results}

Turning now to the results of our GCMC simulations, our strategy for
mapping the coexistence behaviour was initiated on the symmetry plane
$\Delta\mu=0$, where we first mapped the phase diagram in the ($\mu_1,
T$)-plane.  Use of histogram reweighting then provided estimates for
the phase behaviour at small, but finite, $\Delta \mu$.  Guided by
this prediction, a new set of simulations were subsequently performed
at near coexistence state points for this value of $\Delta \mu$, the
results of which were extrapolated to yet larger $\Delta \mu$, and so
on.  In this manner we were able to track the phase behaviour as a
function of $\Delta \mu$ and cover a large range of concentrations in
the process. By accumulating separately contributions to the energy
from like and unlike particle interactions, we were further able to
perform histogram extrapolation with respect to $\delta$.  This was
useful in helping to find the regions of $\delta$ relevant to the two
subtypes.

\begin{figure}[b]
\epsfig{angle=0, file=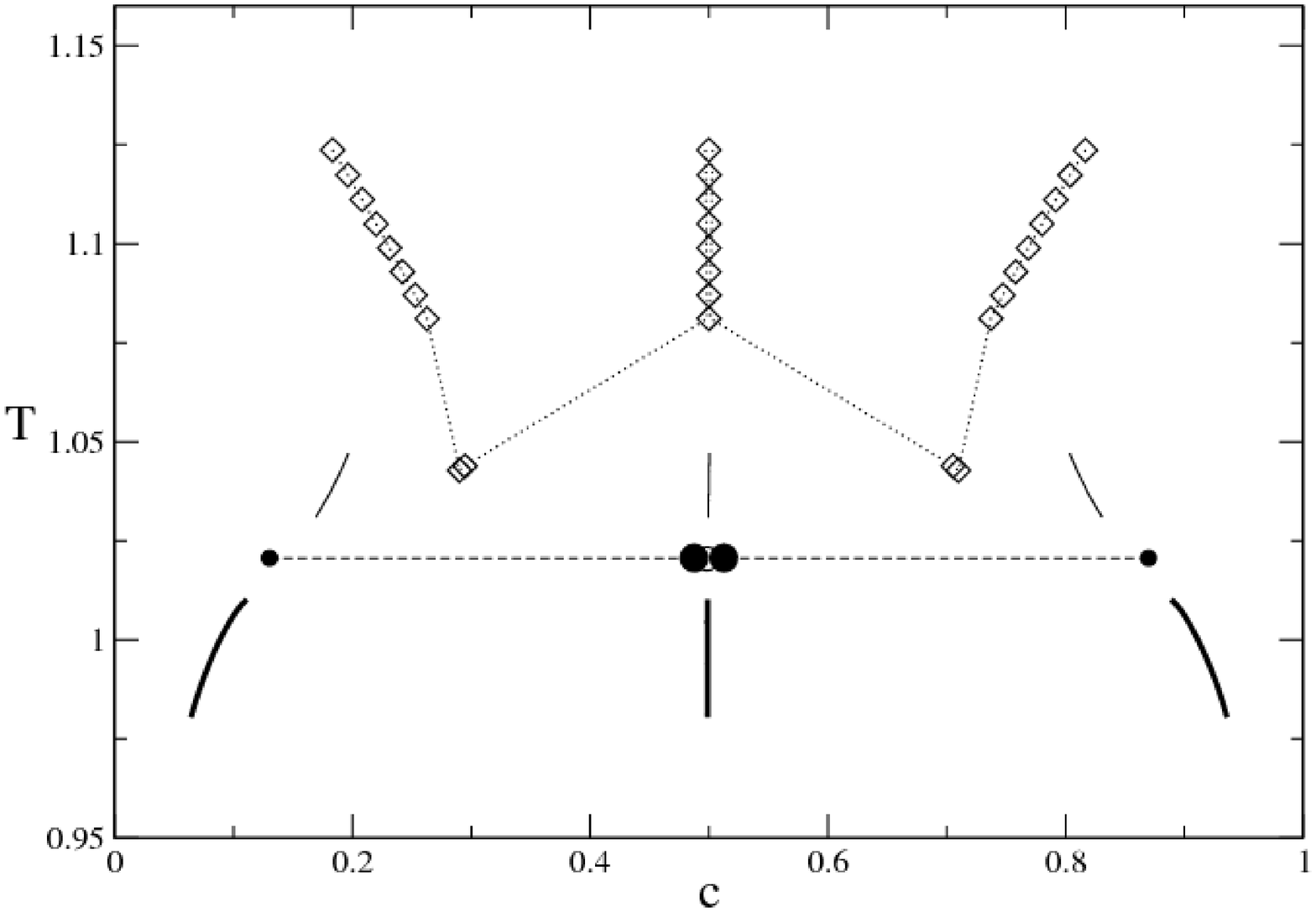, width=6.5cm, clip}
\epsfig{angle=0, file=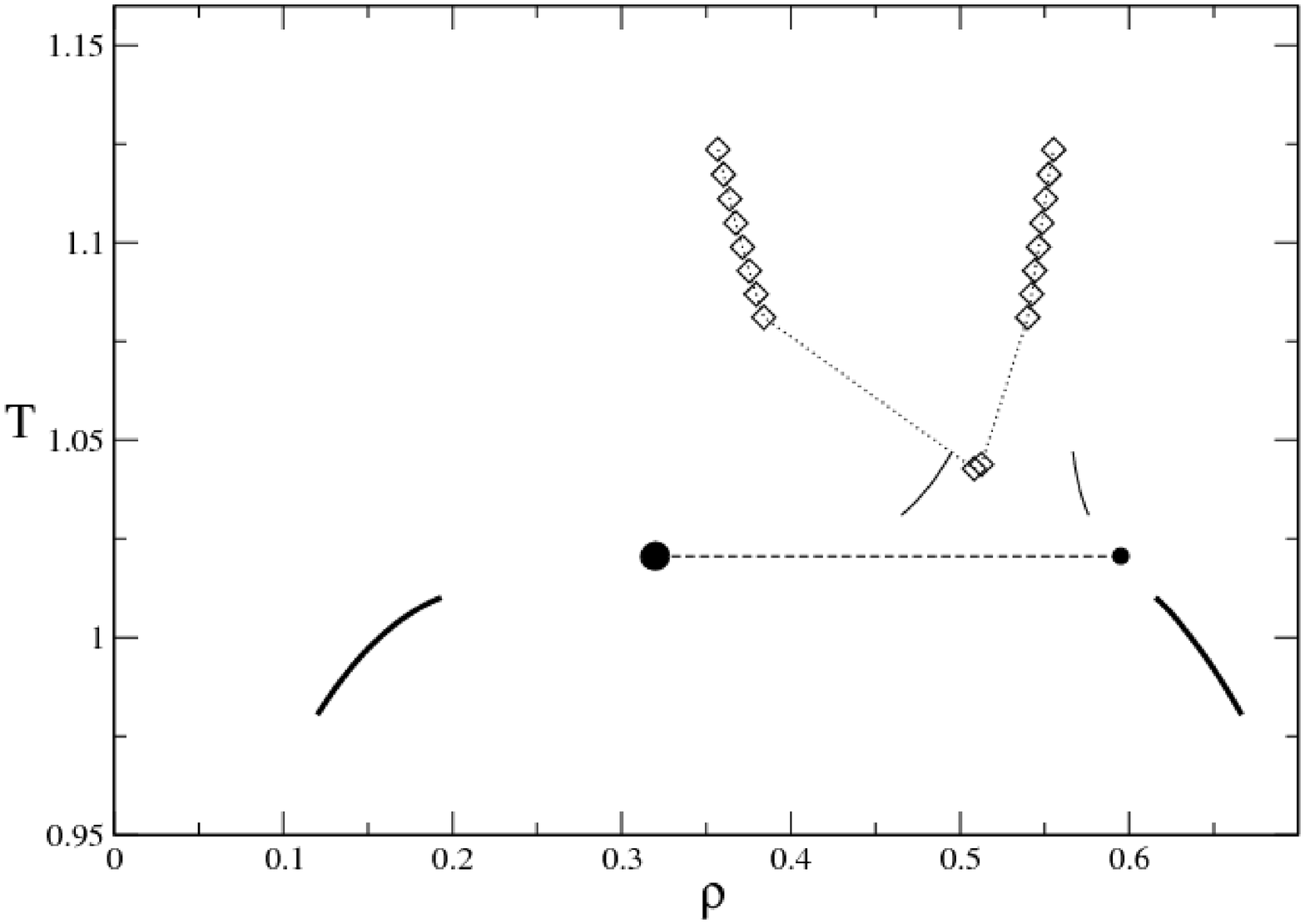, width=6.5cm,clip} 
\epsfig{angle=0, file=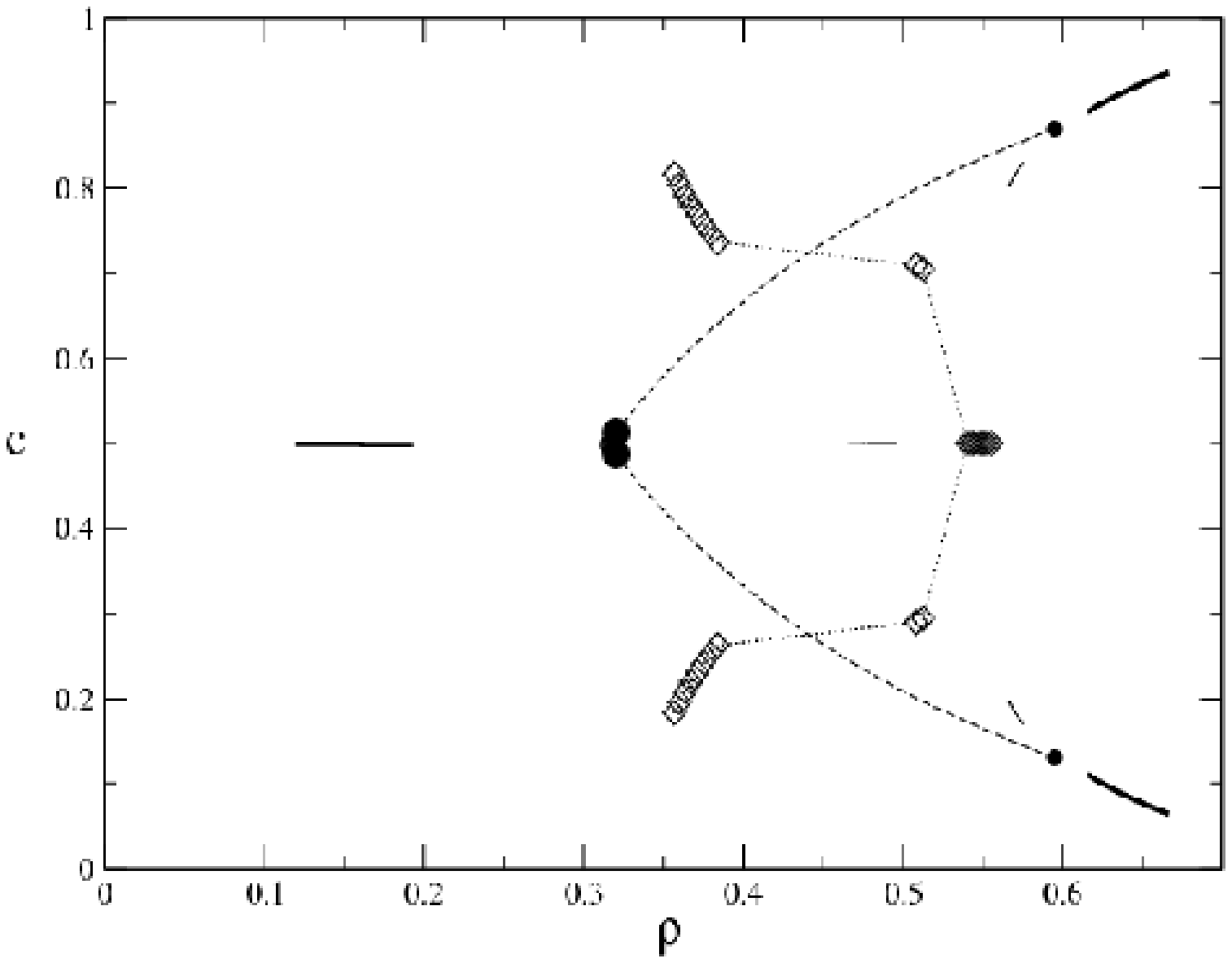, width=6.5cm, clip}
\caption{Projections of the phase diagram measured by simulation for $\delta=0.66$
   (a) $(T, c)$-plane. (b) $(T, \rho)$-plane (c) $(c, \rho)$-plane.
   Solid lines are first order phase boundaries.  Thin solid
   coexistence lines are influenced by finite-size effects and serve
   as a guide to the eye. Diamonds represent points on critical lines
   passing through the field-free tricritical point, whereas circles
   represent points of the critical line passing through the field-free
   LV critical point. Critical points that belong to the same critical
   line are connected via dotted lines.  Large filled symbols
   represent CEPs and small filled symbols the corresponding spectator
   phases.  Tie lines that connect the spectator phase with the CEP
   are shown as dashed lines.} 
\label{fig:MC_0.66proj}
\end{figure}

\begin{figure}[b]
\epsfig{angle=0, file=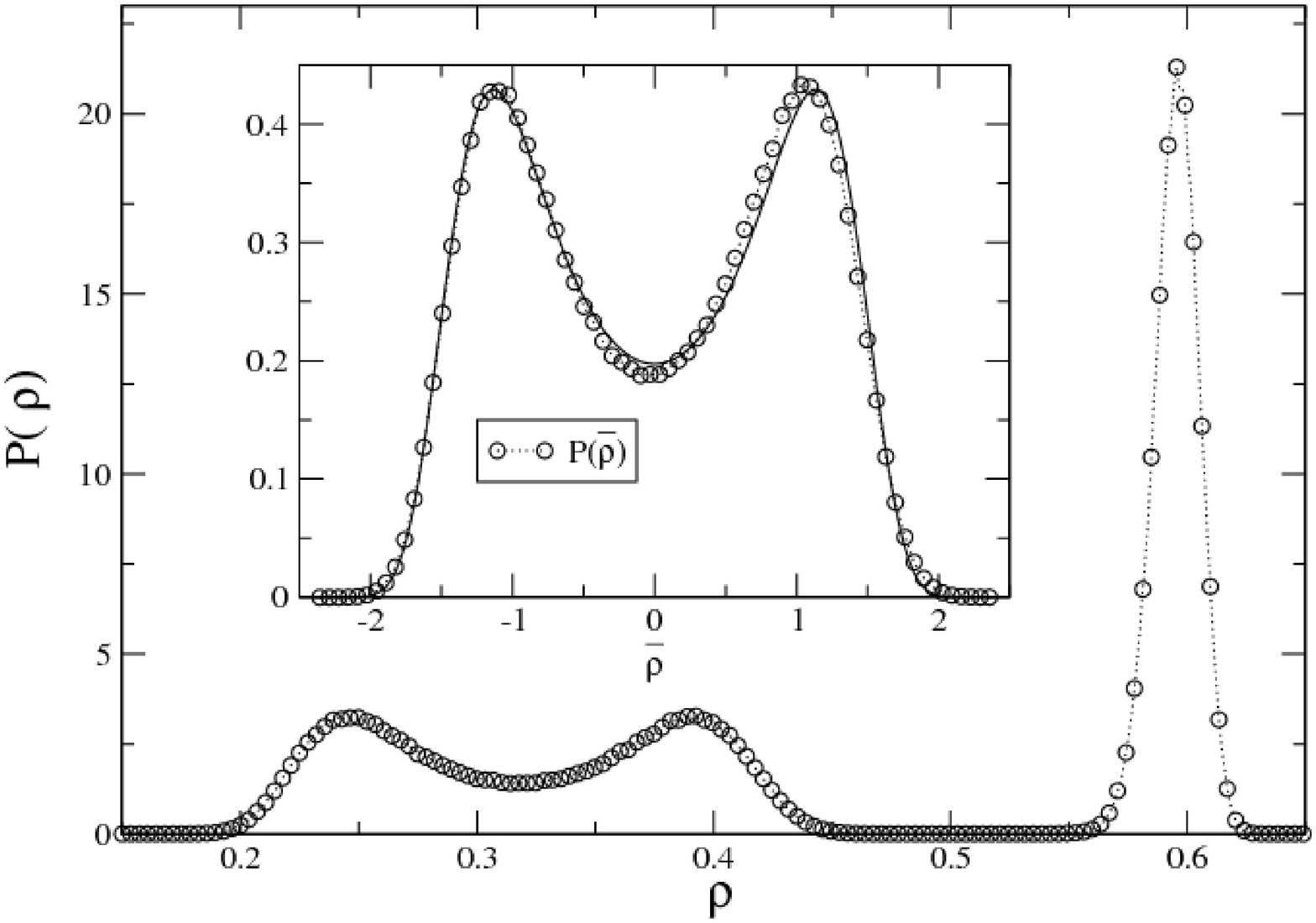, width=6.5cm, clip}
\epsfig{angle=0, file=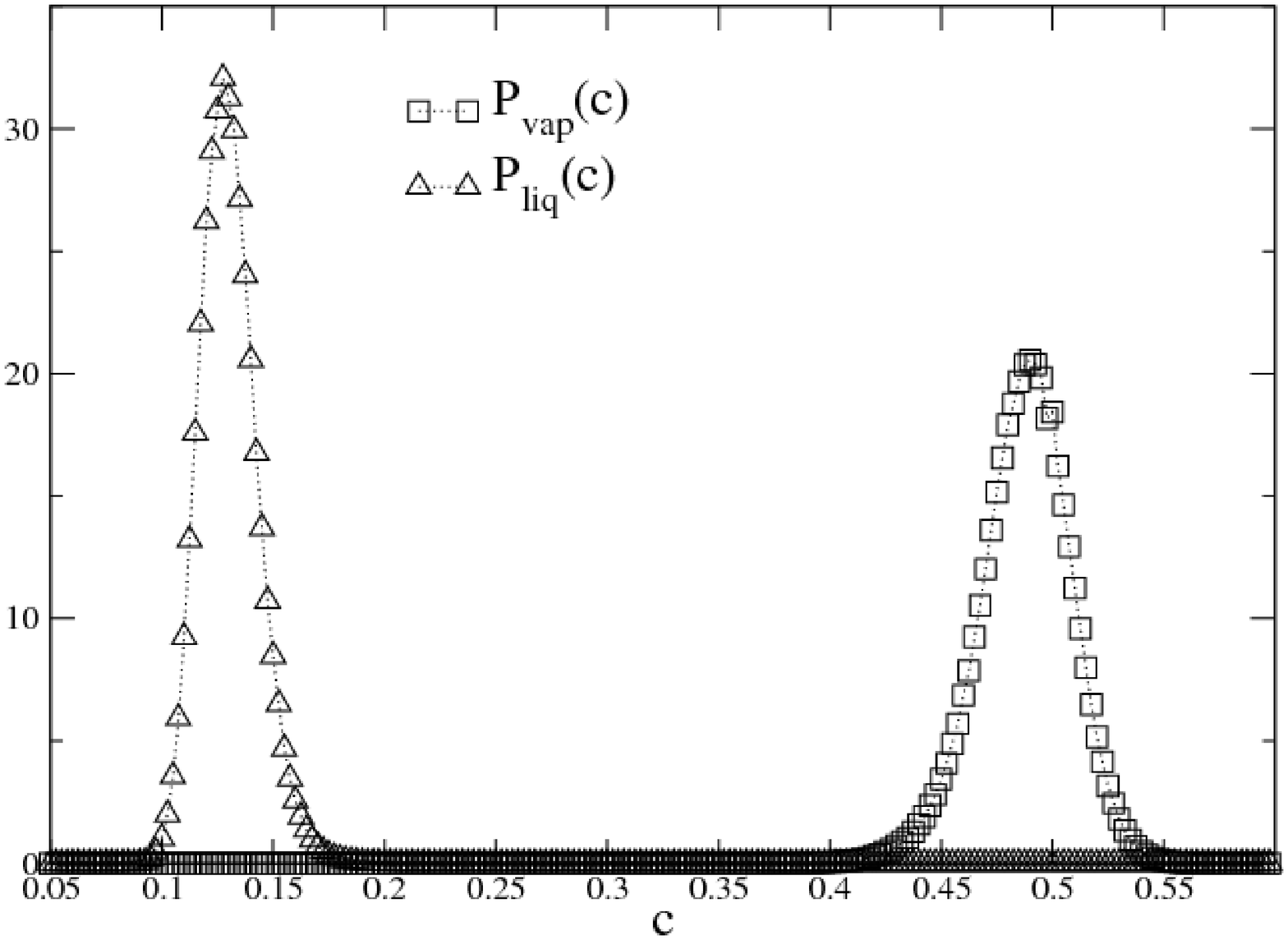, width=6.5cm, clip}
\caption{Critical end point data for $\delta=0.66$. (a) the density
   distribution $P(\rho)$; also shown (inset) is the accord between
   $P(\rho)$ for the critical phase and the appropriately scaled
   universal Ising fixed point form (see text). (b) the concentration
   distribution $P(c)$.}
\label{fig:dists_0.66}
\end{figure}

\begin{figure}[b]
\epsfig{angle=0, file=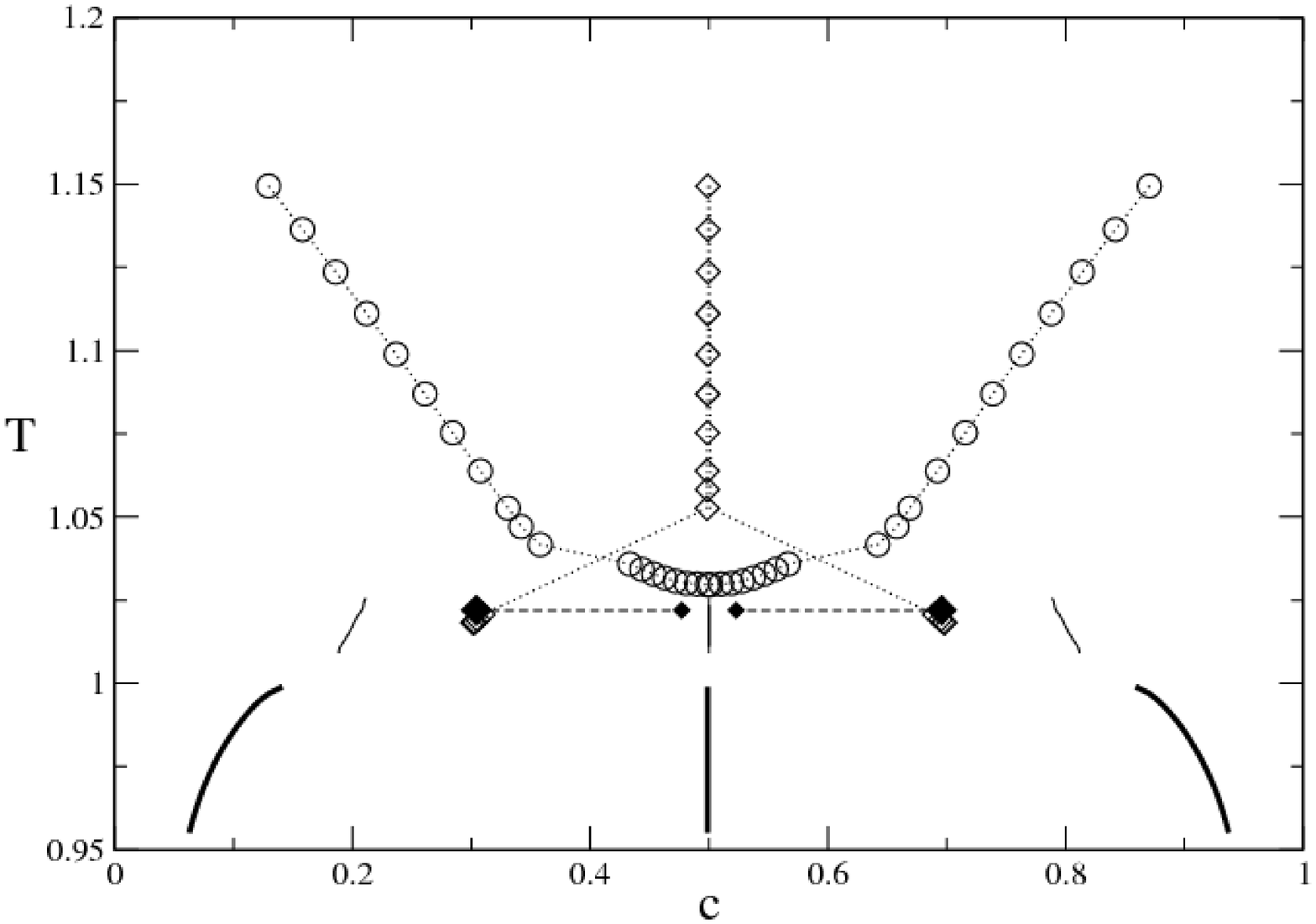, width=6.5cm, clip}
\epsfig{angle=0, file=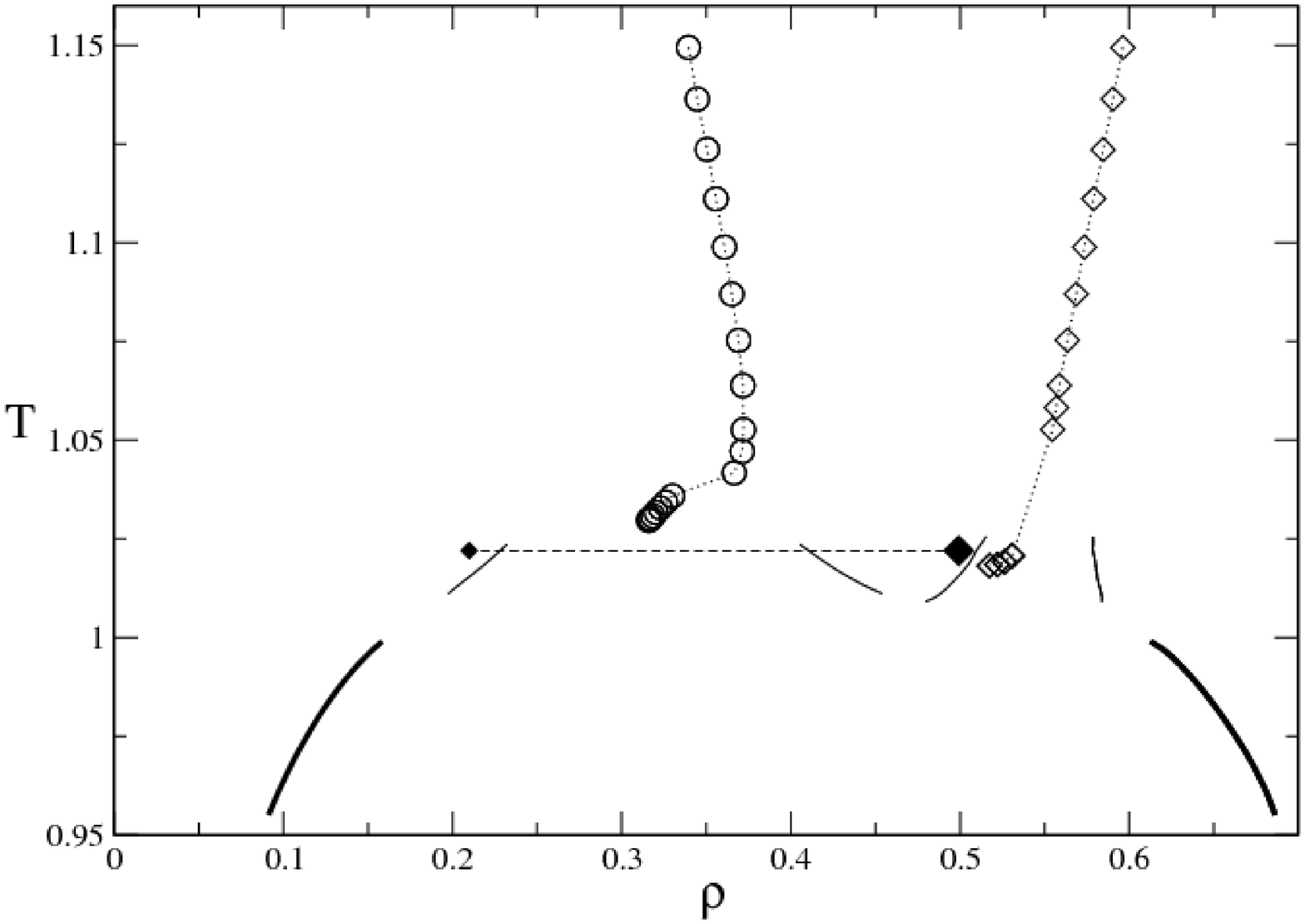, width=6.5cm, clip}
\epsfig{angle=0, file=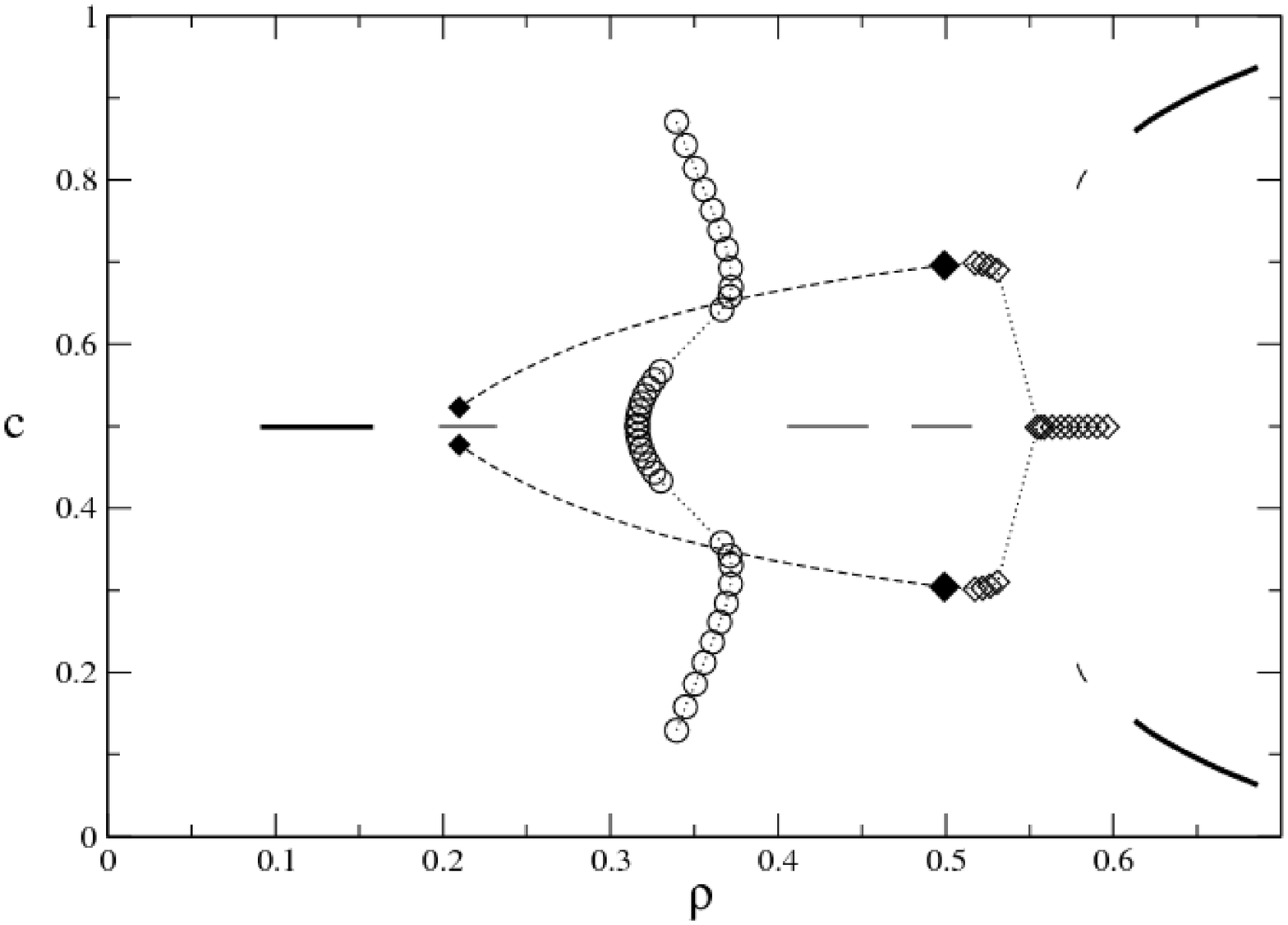, width=6.5cm, clip}
\caption{Projections of the measured phase diagram for $\delta=0.68$
   (a) $(T, c)$-plane, (b) $(T, \rho)$-plane, and (c) $(c,
   \rho)$-plane. For an explanation of symbols, see Fig.~\ref{fig:MC_0.66proj}.}  
\label{fig:MC_0.68proj}
\end{figure}

The high dimensionality of the full phase diagram precludes a
comprehensive study of the full 3d phase diagram as was done via
MSA. Nevertheless projections of our data onto the $(T,\rho)$-,
$(\rho, c)$-, and $(T, c)$-planes, as shown in
Figs.~\ref{fig:MC_0.66proj} and Figs.~\ref{fig:MC_0.68proj}, clearly
confirm the existence of the two subtypes seen in the corresponding
projections of the MSA data (Fig.~\ref{phd_alpha_mixed} and
\ref{phd_beta_mixed}) and detailed above, albeit at slightly different
values of $\delta$. Specifically, for $\delta=0.66$, the critical
lines emanating from the tricritical point links up with that coming
from the LV critical points of the pure phases, while the critical
line emanating from the field-free LV critical point terminates at two
symmetrically related CEPs at which a demixed high density
non-critical (spectator) phase coexists with a lower density critical
phase. The phase that coexist at the CEPs are joined in
Figs.~\ref{fig:MC_0.66proj}(a)-(c) via a dashed line. The character of
the CEP behaviour is clarified by the measured forms of the density
and concentration distributions, as shown in
Fig.~\ref{fig:dists_0.66}. The form of $P(\rho)$ for the critical
phase has a highly non-Gaussian structure which matches the universal
Ising order parameter distribution appropriate for a critical
finite-sized system (see inset) \cite{Wil95,Wil97,Wil03}. That the
critical phase is predominantly liquid-vapor in character is evidenced
by the absence of strong concentrations fluctuations as reflected in
the near-Gaussian form of $P(c)$ for the lower density phase.

\begin{figure}[b]
\epsfig{angle=0, file=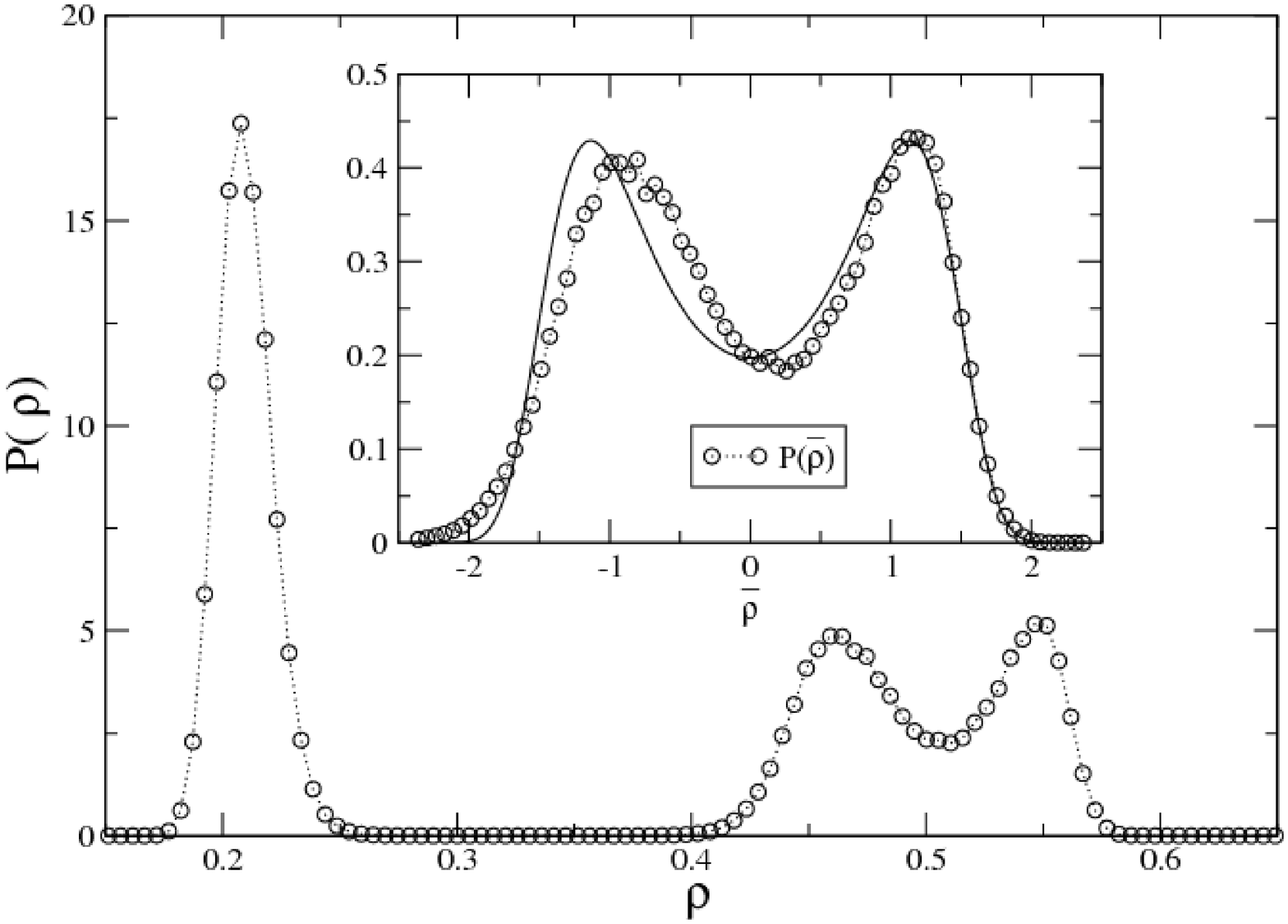, width=6.5cm, clip}
\epsfig{angle=0, file=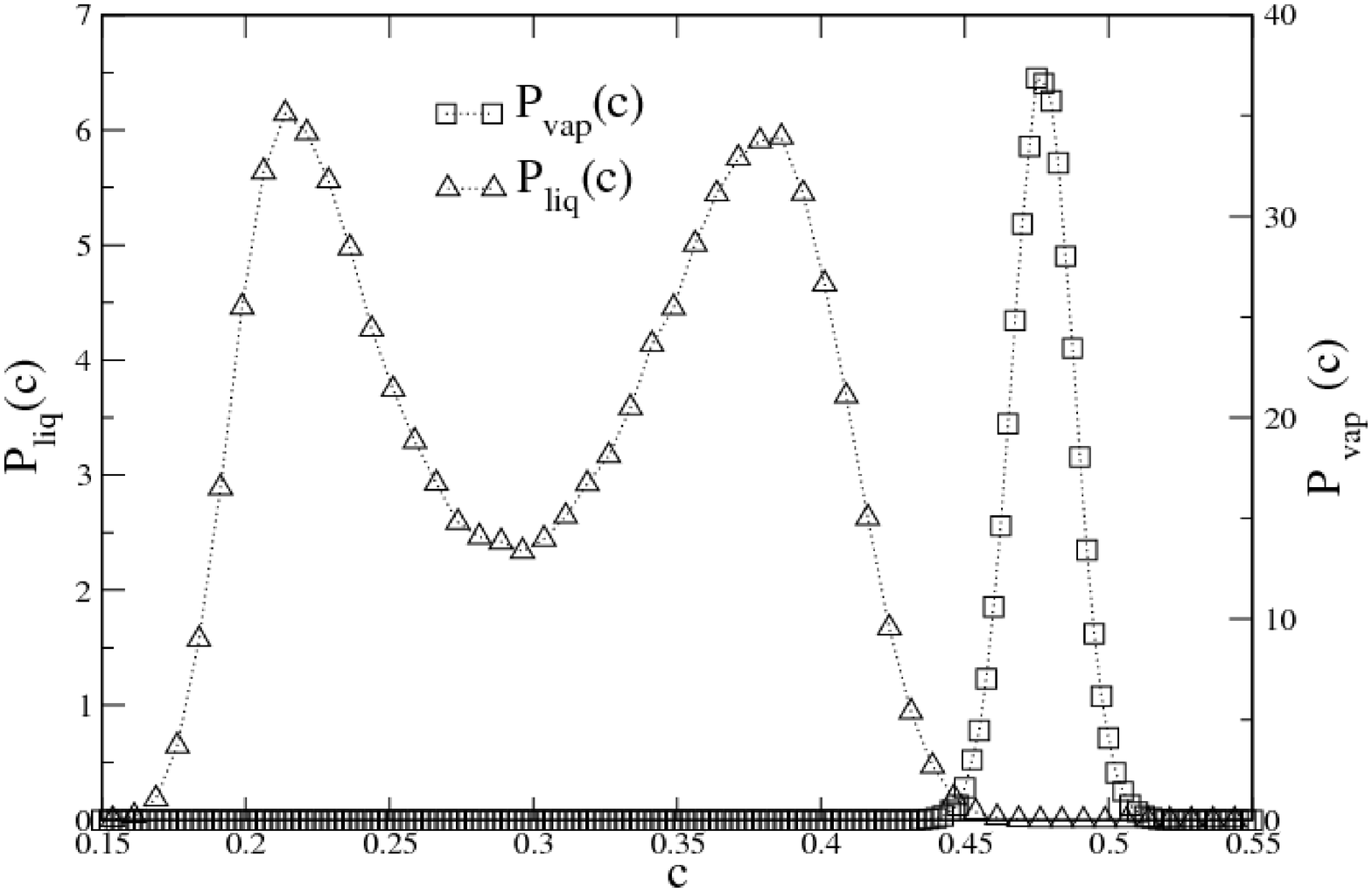, width=6.5cm, clip}
\caption{Critical end point data for $\delta=0.68$. (a) the density
   distribution $P(\rho)$; also shown (inset) is the accord between
   $P(\rho)$ for the critical phase and the appropriately scaled
   universal Ising fixed point form (see text). (b) the concentration
   distribution $P(c)$, where for clarity the two peaks are presented
   on different scales.}
\label{fig:dists_0.68}
\end{figure}

For $\delta=0.68$, the results of Figs.~\ref{fig:MC_0.68proj} shows
that the phase behaviour is of subtype $\beta$. Specifically, the LV
critical point of the pure phases joins smoothly to that of the
equimolar mixture, while the critical line emanating from the
tricritical point terminate at two symmetrically related CEPs, the
coexisting phase of which are joined by dashed lines in
Figs.~\ref{fig:MC_0.68proj}. The nature of these CEPs is again
elucidated by the corresponding forms of $P(\rho)$ and $P(c)$ (see
Fig.~\ref{fig:dists_0.68}). One finds now that the spectator phase is
a low density mixed phase, while the higher density critical phase
exhibits fluctuations that are neither predominantly density-like nor
concentration-like in character, but a coupled mixture of the
two. Thus the forms of $P(\rho)$ and $P(c)$ for the critical phase
both exhibit forms that are highly non-Gaussian in nature, though
neither matches well the universal Ising form. In such situations one
can expect that the true order parameter for this transition involves
a linear combination of density, concentration (and energy), though we
have not attempted to investigate this matter further here.

As regards the range of $\delta$ in which the two subtypes of phase
behaviour occur, the simulation results are in semi-quantitative
agreement with the MSA calculations. However, owing to the
computational burden of this problem, we were not able to pin down
precisely the value of $\tilde\delta$ at which the transition between
subtypes occurs, although our results shows that $\tilde\delta$
differs from that found in MSA by at most $0.01$.

\section{Discussion and conclusions}

\label{sec:concs}

To summarize, we have investigated the full phase behaviour of a
symmetrical binary fluid in a range of model parameters for which, at
equal chemical potentials, the system exhibits a LV critical point and a
tricritical point.  The phase behaviour is considerably richer in both
variety and character than would be expected on the basis of knowledge
of the field-free case alone. Our results confirm previous less detailed
reports that for unequal chemical potentials, two subtypes $\alpha$ and
$\beta$ of phase behaviour occur, and we have elucidated the differences
between these subtypes in terms of the topologies of triple lines and
critical lines.

The MSA results are in semi-quantitative agreement with those of MC
simulations, demonstrating that MSA provides a correct description of
this system. Use of the MSA should therefore prove useful in narrowing
down simulating searches of parameter space when seeking a given type
of phase behaviour. Moreover, our study has demonstrated that
simulations are competitive with theory in providing (within a
reasonable amount of time) precise information on intricate phase
diagrams exhibiting complex topologies of critical lines. Given that
the various commonly used theoretical approaches (HRT, SCOZA, MSA,
$\ldots$) do not always agree, MC thus provides an invaluable
benchmark with which to compare.  We note that our results are in good
agreement with the less comprehensive MF and HRT study of these two
subtypes in ref.~\cite{Pin03}.  Although HRT is generally more
accurate than MSA, it is very laborious to implement and
computationally expensive, and still does not produce results in fully
quantitative agreement with simulation, as has been observed in
studies of the field-free case \cite{Wil03}.

As regards the more general relevance of our findings, one can consider
the symmetrical binary fluid model as probably the simplest member of a
class of one component `spin'-fluid models in which the particles carry
an orientational degree of freedom which features in their interparticle
potentials \cite{Wil98,Pin03}. Other, more complex examples include
Heisenberg \cite{Wei97}, Ising \cite{Hem77}, or dipolar spin fluids
\cite{Gro94}. It is well known that similar sequences of phase diagram
topologies can arise in all members of this class and that one-to-one
correspondences can generally be made between the phases of the binary
symmetrical mixture and those exhibited by the more complex models.
However studies of phase behaviour for the more complex class members
have to be performed using more complicated techniques than the MSA.
Thus the symmetrical mixture plays a key (computationally tractable)
role in elucidating generic aspects of the phase behaviour.

As regards future work, it would be of interest to extend the present
studies to encompass the more general case of asymmetrical
mixtures. Steps in this direction, have recently been reported for the
case of a lattice based binary fluid model \cite{Woy06}, though this
is unable to represent particle species of unequal sizes, whose
packing effects are likely to be subtle.  From the simulation point of
view, GCMC simulations of asymmetrical mixtures are not significantly
more challenging than the symmetrical case, provided the size
asymmetry is moderate. Recent methods for minimizing finite-size
effects in GCMC measurements of coexistence properties of fluid
mixtures, should also help render this a practical proposition
\cite{Buzz06}.

\appendix*
\section{Numerical solution of the MSA}
 
We were able to perform all our MSA calculations using {\tt
MATHEMATICA}\texttrademark~\cite{Math} in a rather straightforward
fashion. Phase diagrams of {\it general} (i.e., non-symmetrical) binary
mixture are also accessible via this numerical route.

We start with the determination of the phase diagram in $(T, \rho,
c)$-space.  To determine at a given temperature $T$, the concentrations
and densities of two coexisting phases (labeled I and II) one
solves the following (coexistence) equations numerically, e.g. with a
Newton-Raphson algorithm

\begin{eqnarray}
\label{MU1} 
\label{COEX} 
\Delta \mu_1^{({\rm I-II})}({\bf z}) & = & 0\\
\label{MU2}
\Delta \mu_2^{({\rm I-II})}({\bf z}) & = & 0\\ 
\label{PRESS}
\Delta p^{({\rm I-II})}({\bf z}) & = & 0, 
\end{eqnarray}
here ${\bf z}=\{ c_{\rm I}, c_{\rm II}, \rho_{\rm I}, \rho_{\rm II}; T
\}$. The $\Delta \mu_i^{({\rm I-II})}$, $i=1,2$, are the differences
between the chemical potentials of the respective species of the
coexisting phases and $\Delta p^{({\rm I-II})}$ is the difference
between the corresponding pressures.

At fixed $T$, equations~(\ref{MU1}) to (\ref{PRESS}) represent three
relations between four unknown quantities, i.e., $c_{\rm I}, c_{\rm II},
\rho_{\rm I}$, and $\rho_{\rm II}$.  To solve this set of nonlinear
equations one requires an additional constraint, which can contain any
subset of the unknown quantities.  The proper choice of such a
constraint is important since it has a distinct influence on the
efficiency of the calculation of coexistence curves.  The simplest (and
most obvious) restriction is to fix the value of a particular variable
and use it to parameterize the coexistence curve.  For example, if we
wish to calculate, at a given $T$, a LV coexistence line, we may fix the
value of the concentration of the vapour phase, $c_{\rm I}$, and solve
equations (\ref{MU1}) to (\ref{PRESS}) for $c_{\rm II}, \rho_{\rm I}$,
and $\rho_{\rm II}$. In order to proceed to a neighboring pair of
coexisting states, we then add to this solution, ${\bf z}$ say, a set of
parameters $\Delta {\bf z} = \{ \Delta c, 0, 0, 0; 0 \}$, i.e., ${\bf
z}' = {\bf z} + \Delta {\bf z}$, which constitutes a starting point for
solving the coexistence equations for the new pair of coexisting states.
The step size $\Delta c$ in the concentration $c_{\rm I}$ depends on the
form of the coexistence lines. Solution of the coexistence equations
leads to a neighboring pair of coexisting states. Given any two states,
new initial values for coexistence points can be obtained via linear
extrapolation from the previous states, thus facilitating use of a
larger step-size in $\Delta c$.

For a triple line (i.e., for three-point coexistence) we have to
satisfy, in addition to equations (\ref{MU1}) to (\ref{PRESS}), the
relations

\begin{eqnarray}
\label{MU1_TRIP}
\Delta \mu_1^{({\rm II-III})}({\bf z}) &=& 0\\
\label{MU2_TRIP}
\Delta \mu_2^{({\rm II-III})}({\bf z}) &=& 0\nonumber\\ 
\label{PRESS_TRIP}
\Delta p^{({\rm II-III})}({\bf z}) &=& 0 \nonumber\\
\end{eqnarray}
with ${\bf z}$ being now ${\bf z} = \{ c_{\rm I}, c_{\rm II}, c_{\rm
III}, \rho_{\rm I}, \rho_{\rm II}, \rho_{\rm III}; T \}$. For fixed
temperature $T$ we thus have six equations for six unknowns, i.e., no
additional constraint is required.

The above considerations provide a recipe for locating coexisting
states, once a solution is known. An appropriate starting point for the
iteration outlined above is a coexistence point of the pure fluids,
i.e., for $c=0$ or $c=1$, for which either equation (\ref{MU1}) or
(\ref{MU2}) is trivially satisfied. Starting at a rather low
temperature coexisting state of the pure fluid, we employ the above
stepwise approach to calculate a {\it full} isothermal cut. While for
these temperatures the coexistence lines of the vapour phase reach the
symmetric plane at $c=1/2$, this is not the case for the coexistence
line of the fluid phase: they terminate instead at the
high-density branch of the symmetrically related triple lines
$\mathfrak{t}_2$ and $\mathfrak{t}_3$. By gradually increasing the
temperature we can trace this pair of triple lines to higher
temperatures and use them at the same time to determine the isothermal
cuts of the demixing coexistence surface, ${\cal S}_1$. For this
region of the phase diagram, we again take advantage of the symmetry of
the system. The variable ${\bf z}$ is then given by ${\bf z} = \{
c_{\rm II}, c_{\rm III} (= 1-c_{\rm II}), \rho_{\rm II}, \rho_{\rm
III} (=\rho_{\rm II}) \}$. In this case equations (\ref{MU1}) and
(\ref{MU2}) are identical and relation (\ref{PRESS}) is trivially
fulfilled for equal densities. Using coexisting states at a given
density and temperature as starting points, we scan the demixing
surface by gradually increasing the temperature at fixed density.

Particular care has to be paid to the determination of the critical
points which, as a function of temperature, form the critical lines.
Since MSA can become numerically unstable in the vicinity of a critical
point, we have sought to localize critical lines within narrow
intervals. To this end we have calculated coexistence states
characterized by a fixed, narrow `distance' $d$, with $d^2 = (c_{\rm
I}-c_{\rm II})^2+(\rho_{\rm I}-\rho_{\rm II})^2$; for $d$ we have
assumed a value of $0.04$. In this manner a good estimate for the
critical line is obtained from the mean value of the concentrations and
densities of the coexisting phases that fulfill the constraint.
	  
The procedure outlined above permits construction of the four
coexistence surfaces, ${\cal S}^\gamma_i$ ($\gamma = \alpha$ or $\beta$
and $i=1, \dots, 4$). The final shape of the phase diagram is then
obtained from the intersections of these coexistence surfaces; this
leads -- as we have learned from the discussion above -- to the triple
lines. These triple lines, which are a key feature of the phase diagram,
constitute the boundaries of the coexistence surfaces, and truncate
metastable regions of the coexisting surfaces, thus defining the final
shape of the {\it full} phase diagram.

Once the phase diagram is determined in $(T, \rho, c)$-space, we
calculate for coexisting states the corresponding thermodynamic
properties (notably the pressure and the chemical potential) and
determine with these quantities the phase diagram in $(T, p, \Delta
\mu)$-space.

\begin{acknowledgments}
JK and GK acknowledge financial support by the \"Osterreichische
Forschungsfond (FWF) under Project Nos. P15758-N08, P17823-N08 and
P17178-N02, the Hochschuljubil\"aumsstiftung der Stadt Wien under
Project No. 1080/2002, and the Au{\ss}eninstitut der TU
Wien. Additional financial support was provided by the Anglo-Austrian
ARC Programme of the British Council. The authors would like to thank
Elisabeth Sch\"oll-Paschinger (Wien) and Davide Pini (Milan) for
useful discussions.
\end{acknowledgments}

\newpage

\begin{table}[htbp]
\centering
\begin{tabular}{|c|c||c|c|}    \hline
\multicolumn{2}{|c||}{van Konynenburg and Scott} & 
\multicolumn{2}{|c|}{Tavares {\it et al.}} \\
\hline
$\Lambda<0$ & I-A   & $\alpha>1$ & no demixing \\ \hline
$\Lambda>0$ &II-A   &$\alpha<$1  & I \\
            &II-A*  &            & II-$\alpha$ \\
            &III-A* &            & II-$\beta$ \\    
            &III-HA &            & III \\
            &not classified &    & IV \\ \hline
\end{tabular}
\caption{Correspondence between the types of binary symmetric mixtures
   classified by van Konynenburg and Scott (see Figs. 1 and 38 of
   \cite{Scott}) and the types I, II, III, and IV introduced by
   Tavares {\it et al.}~\protect\cite{Tav95}.  Types II-$\alpha$ and
   II-$\beta$ are subtypes of type II that are described in more
   detail in~\cite{Koe04}.  $\Lambda$ and $\alpha$ are parameters
   that characterize the relative strength of the mixed interaction of
   the binary (size-)symmetrical mixture. Note that this table
   was already published in \cite{Sch05a}.}
\label{tab:systemparameters}
\end{table}

\newpage

\clearpage

\clearpage

\newpage


\begin{thebibliography}{}
\bibitem{Scott} P.H. van Konynenburg and R.L. Scott, Philos. Trans. R.
		Soc. London, Ser. A {\bf 51}, 495 (1980).
\bibitem{Row82} J.S. Rowlinson and F.L. Swinton, {\it Liquids and Liquid
		Mixtures}, Butterworth (London, 1982), 3rd edition.
\bibitem{Han86} J.-P. Hansen and I.R. McDonald, {\it Theory of Simple Liquids}
		(Cambridge, 2006), 3rd edition.
\bibitem{Fre02} D. Frenkel and B. Smit, {\it Understanding Molecular
		Simulation}, (Academic Press, London, 2002), 2nd edition.
\bibitem{Lan00} D.P. Landau and K. Binder, {\it A guide to Monte Carlo
		simulations in statistical physics}, Cambridge 2000.
\bibitem{Pan88} A.Z. Panagiotopoulos, N. Quirke, M. Stapleton, and D.J. Tildesley,
		Mol. Phys. {\bf 63}, 527 (1988).
\bibitem{Rec93} R.J. Recht and  A.Z. Panagiotopoulos, Mol. Phys. {\bf 80},
		843 (1993).  
\bibitem{Cac93} C. Caccamo and G. Giunta, Mol. Phys. {\bf 78}, 83 (1993).
\bibitem{Gre94} D.G. Green, G. Jackson, E. de Miguel, and L.F. Rull, J. Chem. 
		Phys. {\bf 101}, 3190 (1994).
\bibitem{deM95} E. de Miguel, E.M. del R{\'\i}o, and M.M. Telo da Gama, J. 
		Chem. Phys. {\bf 103}, 6188 (1995).
\bibitem{Wil97} N.B. Wilding, Phys. Rev. E {\bf 55}, 6624 (1997); N.B.
		Wilding Phys. Rev. Lett. {\bf 78}, 1488 (1997).
\bibitem{Cac98} C. Caccamo, D. Costa, and G. Pellicane, J. Chem. Phys. {\bf 109},
		4498 (1998).
\bibitem{Wil98} N.B. Wilding, F. Schmid, and P. Nielaba, Phys. Rev. E 
		{\bf 58}, 2201 (1998).
\bibitem{Ant02} O. Antonevych, F. Forstmann, and E. Diaz-Herrera, Phys. Rev. E 
		{\bf 65}, 061504 (2002).
\bibitem{Kah01} G. Kahl, E. Sch\"oll-Paschinger, and A. Lang, Chemical Monthly 
		{\bf 132}, 1413 (2001).
\bibitem{Wil03} N.B. Wilding, Phys. Rev. E {\bf 67}, 052503-1 (2003).
\bibitem{Woy03} D. Woywod and M. Schoen, Phys. Rev. E {\bf 67}, 026122
		(2003).
\bibitem{Sch03} E. Sch\"oll-Paschinger and G. Kahl, J. Chem. Phys. {\bf 118}, 
		7414 (2003). 
\bibitem{Pin03} D. Pini, M. Tau, A. Parola, and L. Reatto, Phys. Rev. E 
		{\bf 67}, 046116-1 (2003).
\bibitem{Sch04} E. Sch\"oll-Paschinger, E. Gutlederer, and G. Kahl, J. Molec. 
		Liquids {\bf 112},  5 (2004).
\bibitem{Sch05} E. Sch\"oll-Paschinger, D. Levesque, J.-J. Weis, and G. Kahl, 
		J. Chem. Phys. {\bf 122}, 024507-1 (2005).
\bibitem{Sch05a}E. Sch\"oll-Paschinger and G. Kahl, J. Chem. Phys. {\bf 123},
		134508-1 (2005).
\bibitem{Pas01} E. Paschinger, D. Levesque, G. Kahl, and J.-J. Weis, Europhys.
		Lett. {\bf 55}, 178 (2001); E. Sch\"oll-Paschinger, D. 
		Levesque, J.-J. Weis, and G. Kahl, Phys. Rev. E {\bf 64}, 
		011502 (2001).
\bibitem{Woy06} D. Woywod and M. Schoen, Phys. Rev. E {\bf 73}, 011201 (2006).
\bibitem{Par95} A. Parola and L. Reatto, Adv. Phys. {\bf 44}, 211
		(1995); A. Parola and L. Reatto, Phys. Rev. A {\bf 44}, 6600
		(1991); for a more recent overview see also: \cite{Pin03} and 
		\cite{Rei02}.
\bibitem{Rei02} A. Reiner and G. Kahl, Phys. Rev. E {\bf 65}, 046701-1 (2002); 
		A. Reiner and G. Kahl, J. Chem. Phys. {\bf 117}, 4925 (2002).
\bibitem{Koef06}A brief summary of some aspects of this work has previously
		appeared elsewhere: J. K\"{o}finger, G. Kahl, and N.B. Wilding, 
		Europhys. Lett. {\bf 75}, 234 (2006).
\bibitem{Hoy78} L. Blum and J.S. H{\o}ye, J. Stat. Phys. {\bf 19}, 317 (1978).
\bibitem{Arr87} E. Arrieta, C. J{\c e}drzejek, and K. N. Marsh, J. Chem. Phys.
		{\bf 86}, 3607 (1987); {\it ibid.} {\bf 95}, 6806 (1991).
\bibitem{Math}  {\it Wolfram Research Inc., http://www.wolfram.com}.
\bibitem{Pin98} D. Pini, G. Stell, and R. Dickman, Phys. Rev. E {\bf
		57}, 2862 (1998); D. Pini, G.  Stell, and N.B. Wilding, Mol.
		Phys. {\bf 95}, 483 (1998).
\bibitem{Wil95} N.B. Wilding, Phys. Rev. E, {\bf 52}, 602 (1995).
\bibitem{Sco77} R.L. Scott, J. Chem. Soc. Faraday Trans. II {\bf 73},
		356 (1977).
\bibitem{Lomba} E. Lomba, J.-J. Weis, N. G. Almarza, F. Bresme, and G. Stell,
		Phys. Rev. E {\bf 49} (1994) 5169.
\bibitem{CMSL}  N.F. Carnahan and K.E. Starling, J. Chem. Phys. {\bf 51}, 635 
		(1969).
\bibitem{ALLEN}  M.P. Allen and D.J. Tildesley {\em Computer simulation
of liquids} Oxford University Press (1987).
\bibitem{FERRENBERG} A.M. Ferrenberg and R.H. Swendson, Phys. Rev. Lett. {\bf 63},
		1195 (1989).
\bibitem{BERG}  B.A. Berg and T. Neuhaus, Phys. Rev. Lett. {\bf 68}, 9 (1992).
\bibitem{Wil01} N.B. Wilding. Am. J. Phys. 69, 1147 (2001).
\bibitem{Borgs} C. Borgs and R. Kotecky, Phys. Rev. Lett. {\bf 68}, 1734 (1992).
\bibitem{Bloete} M.~M. Tsypin and H.~W.~J. Bl{\"o}te, Phys. Rev. E {\bf
62},  73  (2000).
\bibitem{TRICRIT} I.D. Lawrie and S. Sarbach, in {\em Phase Transition and Critical Phenomena},
edited by C. Domb and J.L. Lebowitz, Vol. 9 (Academic Press, London) 1984.
\bibitem{Wei97} J.M. Tavares, M.M. Telo da Gama, P.I.C. Teixeira, J.-J. Weis, and
		M.J.P. Nijmeier, Phys. Rev. E {\bf 52}, 1915 (1995); M.J.P. 
		Nijmeijer and J.-J. Weis, Phys. Rev. Lett. {\bf 75}, 2887 (1995);
                M.J.P. Nijmeijer and J.-J. Weis, Phys. Rev. E {\bf 53}, 591 
		(1996); J.-J. Weis, M.J.P. Nijmeijer, J.M. Tavares, and M.M. 
		Telo da Gama, Phys. Rev. E {\bf 55}, 436 (1997).
\bibitem{Hem77} P.C. Hemmer and D. Imbro, Phys. Rev. A {\bf 16}, 380 (1977); 
		W. Fenz, R. Folk, I.M. Mryglod, and I.P. Omelyan, Phys. Rev. E
		{\bf 68}, 061510-1 (2003); I.P. Omelyan, I.M. Mryglod, R.
		Folk, and W. Fenz, Phys. Rev. E {\bf 69}, 061506-1 (2004);
		I.P. Omelyan, W. Fenz, I.M. Mryglod, and R. Folk, Phys. Rev.
		Lett. {\bf 94}, 045701-1 (2005).
\bibitem{Gro94} B. Groh and S. Dietrich, Phys. Rev. E {\bf 50}, 3814 (1994);
		B. Groh and S. Dietrich, Phys. Rev. Lett. {\bf 72}, 2422 
		(1994); {\it ibid.} {\bf 74}, 2617 (1997).
\bibitem{Tav95} J.M. Tavares, M.M. Telo da Gama, P.I.C. Teixeira, J.-J. Weis, 
		and M.J.P. Nijmeijer, Phys. Rev. E {\bf 52}, 1915 (1995).
\bibitem{Buzz06} M Buzzacchi, P. Sollich, N.B. Wilding and M. Müller,
Phys. Rev. E73, 046110 (2006).
\bibitem{Koe04} J. K\"ofinger, Diploma-thesis (Technische Universit\"at Wien,
		2004, unpublished). 
\end{thebibliography}
\end{document}